\documentclass[aps,pra,onecolumn,superscriptaddress]{revtex4-2}

\usepackage[utf8]{inputenc}
\usepackage[normalem]{ulem}
\usepackage{mathtools}
\usepackage{graphicx}
\usepackage{hyperref}
\usepackage{xfrac}

\usepackage{cleveref}
\usepackage{float}
\restylefloat{table}

\usepackage{amsmath,amssymb,amsxtra,amstext,amsfonts,dsfont}
\usepackage{multirow}
\usepackage{array}
\usepackage{upgreek}
\usepackage{mathrsfs}
\usepackage{bbold}
\usepackage{bm}

\usepackage{soul}
\setstcolor{red}

\usepackage{physics}
\usepackage{xcolor}

\newcommand{\xac}{\hat x_A^{(C)}}
\newcommand{\pac}{\hat p_A^{(C)}}
\newcommand{\Hac}{\hat H_A^{(C)}}
\newcommand{\xbc}{\hat x_B^{(C)}}
\newcommand{\pbc}{\hat p_B^{(C)}}
\newcommand{\Hbc}{\hat H_B^{(C)}}

\newcommand{\xca}{\hat x_C^{(A)}}
\newcommand{\pca}{\hat p_C^{(A)}}

\newcommand{\xba}{\hat x_B^{(A)}}
\newcommand{\pba}{\hat p_B^{(A)}}

\newcommand{\Dac}{\hat D_A^{(C)}}

\newcommand{\Cov}{\text{Cov}}

\begin{document}

\title{Trajectory sharpening under quantum reference frame transformations}

\author{Matteo Fadel}
\email{matteofadel@pku.edu.cn}
\affiliation{Institute of Quantum Science and Technology $\&$ Frontiers Science Center for Nano-Optoelectronics, School of Physics, Peking University, Beijing 100871, China}
\affiliation{Department of Physics, ETH Z\"{u}rich, Z\"{u}rich 8093, Switzerland}

\author{Giulia Gubitosi}
\affiliation{Dipartimento di Fisica ``E. Pancini'', Università di Napoli Federico II, Napoli 80126, Italy}
\affiliation{INFN sezione di Napoli, Napoli 80126, Italy}

%\maketitle
%\tableofcontents

\begin{abstract}
Quantum reference frame (QRF) transformations relate the description of a physical system to different quantum observers, but their connection to concrete experimental phenomena remains largely unclear. Here we show that the trajectory-sharpening effect obtained by correlating a probed system with an effective negative-mass reference admits a natural interpretation as a QRF transformation to the frame of a negative-mass system. 
Using a minimal toy model of free particles with signed masses, we identify the Einstein-Podolsky-Rosen-type observables responsible for this effect with a pair of commuting relational observables that, under the instantaneous QRF transformation, are mapped to operators local to distinct subsystems. We then show that the corresponding relational trajectory is reproduced by the dynamical QRF transformation, where the transformed relational position acquires an explicit time dependence.
We introduce QRF-invariant sharpness measures for relational trajectories and derive monogamy relations showing that a trajectory cannot be made simultaneously arbitrarily sharp with respect to two distinct negative-mass frames.
We further show that trajectory sharpness can instead reside in a collective reference mode encoded across two physical systems, even when neither constituent alone provides an arbitrarily sharp reference, so that sharpness can be distributed rather than localized in a single constituent frame. These results provide an operational, experimentally relevant interpretation of QRF transformations for continuous-variable systems.
\end{abstract}

\maketitle

%%%%%%%%%%%%%%%%%%%%%%%%%%%%%%%%%%%%%%%%%%%%%%%%%%%%%%%%%%%%%%%%%%%%%

\section{Introduction}

Quantum reference frames (QRFs) have been proposed to describe reference frames associated with quantum systems~\cite{Aharonov:1984zz, Palmer:2013zza, Vanrietvelde:2018pgb}. They are especially relevant in quantum information~\cite{Peres:2002wx, Bartlett2007} and quantum gravity~\cite{Rovelli:1990pi, Vanrietvelde:2018pgb}, where the quantum nature of reference systems can fundamentally affect the description of physical phenomena. 
The formalism developed in~\cite{Giacomini:2017zju} uses tools of relational quantum mechanics to describe physical systems whose states are quantum relative to one another, and to define transformations between them. In particular, when the relative state of quantum systems is described by a superposition of positions and velocities, the transformation between their respective frames depends on the quantum state of the frames themselves. 
These transformations provide a quantum extension of the covariance principle and can be understood as symmetries of the underlying physical systems~\cite{Giacomini:2017zju, Krumm:2020fws, delaHamette:2020dyi, Apadula:2022pxk}, forming a non-trivial group structure ~\cite{Ballesteros:2020lgl,Ballesteros:2025ypr}.
Additionally, this formalism provides a framework for studying how the description of superpositions, entanglement, and uncertainty relations depend on the choice of quantum reference frame~\cite{Palmer:2013zza, Giacomini:2017zju, Brukner:2018oda,Riera:2024ehk}, and has been applied to the study of indefinite causal structure in quantum gravity~\cite{Castro-Ruiz:2017gzs, Castro-Ruiz:2019nnl, delaHamette:2021iwx}. Despite these theoretical developments, a direct connection between the abstract unitary transformations relating different QRFs and concrete, measurable laboratory phenomena is still largely
missing.

A seemingly unrelated line of work has shown that Einstein-Podolsky-Rosen (EPR) correlations~\cite{einstein_can_1935} between two massive systems can be exploited to track the trajectory of one of them without the quantum back-action penalty that would otherwise be imposed by the Heisenberg uncertainty relation, provided that one of the two systems behaves as an effective negative-mass oscillator~\cite{tsang_evading_2012,Polzik:2014xza}. 
In that scheme, the relative position and the total momentum of the pair are commuting observables (with the total momentum conserved), so that the trajectory of one system relative to the other can in principle be tracked with arbitrary precision. 
This scheme was originally introduced as a metrological tool for force sensing beyond the standard quantum limit, and has been realized experimentally \cite{moller_quantum_2017}.

In this work, we develop a direct interpretation of the negative-mass construction in terms of quantum reference frames: it amounts to describing the probed system relative to a quantum reference system whose effective negative mass allows the relative trajectory to become sharply defined. We make this connection explicit and use it to reinterpret and extend the trajectory-sharpening effect of the EPR scenario within the operational QRF framework.

Starting from a minimal toy model of two free particles with opposite-sign masses, we show that the EPR-correlated observables are precisely the commuting relational observables that, under the instantaneous position-QRF transformation, are mapped to operators local to distinct subsystems in the QRF of the negative-mass particle. We then show that the same relational trajectory is reproduced by the dynamical QRF transformation.
We clarify how the dynamical description differs between the time-dependent and fixed-time (initial-slice)  transformations, and show that, although they distribute the time dependence differently between the transformed Hamiltonian and observables, they reproduce the same relational trajectory.

Building on this, we introduce QRF-invariant measures of trajectory sharpness based on the uncertainty properties of the relevant pair of commuting relational observables. We consider both a marginal uncertainty product and a covariance-area measure that also accounts for correlations between the two observables. Both are QRF-invariant, and the latter is exactly conserved under the opposite-mass free evolution. 
These quantities reveal a monogamy relation: making the trajectory of a probe system arbitrarily sharp relative to one negative-mass reference frame necessarily limits how sharp the same trajectory can be relative to another distinct frame.
Moreover, the covariance-area formulation provides a stronger version of this trade-off, showing that the restriction persists even after correlations within each relational pair are fully taken into account.

Finally, we show that this restriction on simultaneous sharpness with respect to the two local frames does not prevent sharpness from residing in a collective reference degree of freedom. A suitably chosen collective mode encoded across two physical systems can define an arbitrarily sharp relational frame for $B$, even when, for the class of states considered, neither constituent system alone can do so. This shows that trajectory sharpness can be encoded in a distributed collective mode rather than localized in a single constituent reference frame.

These results provide an experimentally grounded interpretation of QRF transformations for continuous-variable systems, and suggest concrete physical platforms, e.g.\ optomechanical or atomic-ensemble implementations of a negative-mass reference, for probing the genuinely quantum features of reference-frame changes.

%%%%%%%%%%%%%%%%%%%%%%%%%%%%%%%%%%%%%%%%%%%%%%%%%%%%%%%%%%%%%%%%%%%%%
\section{Toy model from the Einstein-Podolsky-Rosen paradox}
\label{sec:toymod}

In order to better understand the main ideas behind our work, let us start by presenting the following illustrative model.
Consider three systems, labeled $A$, $B$ and $C$, with masses $m_A$, $m_B$ and $m_C$, respectively. Taking $C$ as the laboratory reference system, $\xbc$, $\pbc$ are the position and momentum operators for $B$.
These satisfy the canonical commutation relation $[\xbc,\pbc]=i\hbar$, which implies the Heisenberg uncertainty relation $\Delta\xbc \Delta \pbc \geq \hbar/2$. 
This inequality is known to pose a limitation on the precision with which non-commuting observables can be simultaneously known.
For example, if at a given time the position of $B$ is precisely known, meaning a standard deviation $\Delta\xbc\rightarrow 0$, then this must come at the expense of having large momentum uncertainty $\Delta\pbc\rightarrow \infty$.

Taking $B$ to be a free particle governed by the Hamiltonian $\Hbc=(\pbc)^2/2m_B$, in the Heisenberg picture the time evolution of its position is 
\begin{equation}\label{eq:EOMxbc}
    \xbc(t) = \xbc(0) + \dfrac{\pbc(0)}{m_B} t \;.
\end{equation}

\begin{figure}[!t]
    \centering
    \includegraphics[width=\linewidth]{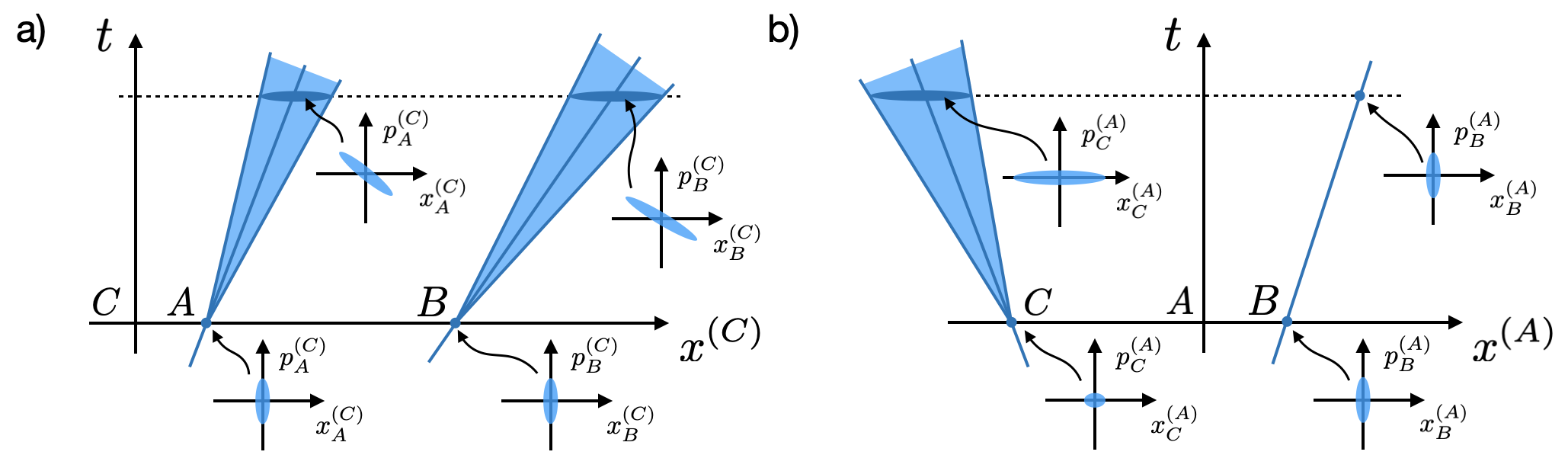}
    \caption{\textbf{Illustration of a reference frame transformation between three particles.} 
    a) In the reference frame of $C$, systems $A$, $B$ can have well-defined position at time $t=0$, but at the cost of giving trajectories with position uncertainty for $t>0$ (blue shaded regions).
    b) Relative to $A$, however, the trajectory of system $B$ can be made arbitrarily sharp at any fixed time, or uniformly over any prescribed finite time interval, by simultaneously reducing the uncertainties of the commuting EPR observables. In the ideal joint-sharpness limit the entire relational trajectory is sharply defined. Its time evolution is governed by $\pca$, rather than by $\pba$, see Eq.~\eqref{eq:xB_A_eq35}.
    }
    \label{fig:1}
\end{figure}

This implies non-trivial commutation relations for the position along the trajectory 
\begin{equation}\label{eq:CommDifftX}
    [\xbc(0),\xbc(t)] = i\hbar \dfrac{t}{m_B} \;.
\end{equation}
Therefore, in the laboratory frame of reference, the position of an ordinary free particle cannot be arbitrarily sharp at two distinct times, see Fig.~\ref{fig:1}a:
\begin{equation}\label{eq:CommDifftX1}
  \Delta\xbc(0)\Delta\xbc(t) \geq \dfrac{\hbar}{2} \dfrac{|t|}{|m_B|}  \;.
\end{equation}

However, the trajectory can be made sharp with respect to another system. To see this, let us consider system $A$, whose quadratures $\xac$ and $\pac$ in the laboratory frame  satisfy the commutation relation $[\xac,\pac]=i \hbar$, and whose dynamics is described by the free Hamiltonian $\Hac=(\pac)^2/2m_A$. The time evolution of its position is given by
\begin{equation}\label{eq:EOMxac}
    \xac(t) = \xac(0) +  \dfrac{\pac(0)}{m_A} t \;.
\end{equation}

Eqs.~(\ref{eq:EOMxbc}) and (\ref{eq:EOMxac}) can be combined to yield  the time evolution of the relative position between $A$ and $B$, as observed from the laboratory $C$,
\begin{equation}
\xbc(t)-\xac(t) = \xbc(0)-\xac(0) + \left(\dfrac{\pbc(0)}{m_B} - \dfrac{\pac(0)}{m_A}\right)t .
\end{equation}
Assuming that $m_B=m>0$ and $m_A=-m$, such that system $A$ has negative mass, we then obtain
\begin{equation}\label{eq:EOMrelX}
\xbc(t)-\xac(t) = \xbc(0)-\xac(0) + \dfrac{1}{m}\left(\pbc(0) + \pac(0)\right)t \;.
\end{equation}
Crucially, as originally pointed out by Einstein, Podolsky, and Rosen \cite{einstein_can_1935}, these observables satisfy
\begin{equation}
[\xbc(0) - \xac(0), \pbc(0) + \pac(0)] = 0 \;,
\end{equation}
so that these two linear combinations of quadratures  can, in principle, be known simultaneously with arbitrary precision. Consequently, one can predict the relative position $\xbc(t)-\xac(t)$ at any later time with arbitrary precision, as illustrated in Fig.~\ref{fig:1}b. In fact, note that contrary to Eq.~\eqref{eq:CommDifftX} we now have
\begin{equation}
    [\xbc(0) - \xac(0), \xbc(t) - \xac(t)] = 0 \;.
\end{equation}

One might wonder about the concrete relevance of this argument, since a system with negative mass may seem unphysical. 
However, effective negative-mass systems have already been realized within existing experimental platforms.
The paradigmatic example is a collective atomic spin ensemble polarized (anti-)parallel to a static magnetic field: for a spin oriented along the field, small transverse fluctuations behave as a harmonic oscillator with positive effective mass, whereas for a spin oriented against the field the same fluctuations are described by an oscillator with effective mass of opposite sign~\cite{Polzik:2014xza}. 
By combining such a negative-mass spin oscillator with an ordinary, positive-mass system, such as another atomic ensemble, or a
mechanical or optical mode, and coupling them together (e.g. through a light-mediated quantum-non-demolition or beam splitter-type interaction), one can realize the corresponding EPR-type situation in which suitable relative-position and collective-momentum quadratures are simultaneously squeezed below the standard quantum limit, as demonstrated experimentally in atomic ensembles \cite{wasilewski_quantum_2010,julsgaard_experimental_2001,Fadel:2018krh,kunkel_spatially_2018} and mechanical platforms~\cite{moller_quantum_2017,thomas_entanglement_2021,ockeloen-korppi_stabilized_2018,mercier_de_lepinay_quantum_2021,kotler_direct_2021,riedinger_remote_2018,kern_einstein-podolsky-rosen_2026}. 

In a broader picture, we can understand this toy example as a transformation of reference frames or, more precisely, quantum reference frames.
In particular, Eq.~\eqref{eq:EOMxbc} shows that the trajectory of system $B$, as seen from $C$, cannot show arbitrarily small uncertainty at all times. 
On the other hand, Eq.~\eqref{eq:EOMrelX} tells us that the relative position between $B$ and $A$ can in principle be made arbitrarily sharp at any fixed time, or over any prescribed finite time interval, and becomes sharp at all times in the ideal joint-sharpness limit of the two commuting EPR observables.
Our goal is to formalize these observations in the language of quantum reference frames to explore possible implications.

%%%%%%%%%%%%%%%%%%%%%%%%%%%%%%%%%%%%%%%%%%%%%%%%%%%%%%%%%%%%%%%%%%%%%
\section{Quantum reference frame transformations and dynamical evolution}

In this Section, we briefly review QRF transformations introduced in~\cite{Giacomini:2017zju}, following the notation presented in Refs.~\cite{Ballesteros:2020lgl, Ballesteros:2025ypr}. We restrict the discussion to results that are directly relevant to the scope of this work. 
In particular, to establish the link with the physical setup described in the previous Section, we consider time-dependent translations to the QRF of a system whose dynamics is governed by a free Hamiltonian. 
To account for cases where relative velocities of the systems at play may be relevant,  one can also consider QRF boosts~\cite{Giacomini:2017zju, Giacomini:2018gxh,Ballesteros:2020lgl}. Since boosts depend on the mass of the system, extending them to the negative-mass case requires some care, as discussed in Appendix~\ref{sec:supp-boost}.

\subsection{QRF transformations}
\label{sec:QRFdefAndGroup}

The simplest QRF transformation describes the transformation of  the phase space coordinates of a quantum system relative to a quantum particle $C$ into the phase space coordinates relative to another quantum particle $A$ at a fixed time. 
Specifically, the transformation of phase space coordinates $(\xac,\pac,\xbc,\pbc)$ of particles $A$ and $B$ relative to particle $C$ to the phase space coordinates $(\xca,\pca,\xba,\pba)$ of particles $C$ and $B$ relative to $A$ is implemented by the unitary operator
\begin{equation}
\hat S_x^{(C \rightarrow A)}= \hat{\mathcal P}_{AC}\, e^{\frac{i}{\hbar} \xac \pbc},
\label{eq:Sx_def}
\end{equation}
with \(\hat{\mathcal P}_{AC}\) the parity-swap operator between the old frame $C$
and the new frame $A$, acting on $A$ as $\hat{\mathcal{P}}_{AC}\,\xac\, \hat{\mathcal{P}}_{AC}^\dagger = -\xca$, $ \hat{\mathcal{P}}_{AC}\,\pac\, \hat{\mathcal{P}}_{AC}^\dagger = -\pca$, and on $B$ as $ \hat{\mathcal{P}}_{AC}\,\xbc\, \hat{\mathcal{P}}_{AC}^\dagger = \xba$, $ \hat{\mathcal{P}}_{AC}\,\pbc\, \hat{\mathcal{P}}_{AC}^\dagger = \pba$.

Explicitly, the QRF transformation $\hat S_x^{(C\to A)}$ acts as
(omitting here the superscript $(C\to A)$ for simplicity)
\begin{subequations}\label{eqs:SxDef}
\begin{align}
\hat S_x \, \xac \, \hat S_x^\dagger &= -\xca, & \hat S_x \, \pac \, \hat S_x^\dagger &= -\pca - \pba, \\
\hat S_x \, \xbc \, \hat S_x^\dagger &= \xba - \xca, & \hat S_x \, \pbc \, \hat S_x^\dagger &= \pba .
\end{align}
\end{subequations}
This transformation maps the position and momentum operators of $A$ and $B$ relative to $C$ into the position and momentum operators of $B$ and $C$ relative to $A$. 
Note that the canonical commutation relations $[\xac,\pac]=[\xbc,\pbc]=i\hbar$ are preserved under the transformation, as expected for a canonical transformation.
Crucially, let us emphasize that this position-based QRF change from frame $C$ to frame $A$ is purely kinematical and does not depend on the masses. 

Taking into account the dynamical evolution of the frames $A$ and $C$, one can further generalize the previous transformations to define generalized Galilean translations. When $A$ and $C$ evolve according to a free-particle Hamiltonian, these transformations are generated by the unitary operator
\begin{align}\label{eq:STSb}
\hat S_T^{(C\to A)}
&=
e^{-\frac{i}{\hbar}\hat Q_C^{(A)} t}\,
\hat{\mathcal{P}}_{AC}\,
e^{\frac{i}{\hbar}\hat P_{AB}^{(C)}}\,
e^{\frac{i}{\hbar}\hat Q_A^{(C)} t},
\end{align}
where $\hat Q_i^{(j)} = \bigl(\hat p_i^{(j)}\bigr)^2 / 2 m_i$ is the free-particle Hamiltonian of particle $i$ from the point of view of $j$ and $\hat P_{AB}^{(C)} = \xac \pbc$.
The corresponding action on phase space coordinates is (again omitting the superscript $(C\to A)$)
\begin{subequations}\label{eq:DefSTtransf}
\begin{align}
\hat S_T\,\xac\,\hat S_T^\dagger &= -\xca + \pca\, t\left(\frac{1}{m_C}-\frac{1}{m_A}\right) - \frac{\pba}{m_A}\,t, \\
\hat S_T\,\pac\,\hat S_T^\dagger &= -\pca-\pba, \\
\hat S_T\,\xbc\,\hat S_T^\dagger &= \xba-\xca+\frac{\pca}{m_C}\,t, \\
\hat S_T\,\pbc\,\hat S_T^\dagger &= \pba. 
\end{align}
\end{subequations}
The physical meaning of this transformation is that the position of system
$B$ at time $t$ in the frame of $C$ is mapped to the relative
position between system $B$ at time $t$ and system $C$ at time $t=0$ in the frame of $A$. Note that the momentum of $B$ remains unchanged, as expected for a translation.

These transformations satisfy the transitivity property, meaning that the change of reference frame from $C$ to $A$ is equivalent to the change from $C$ to an intermediate frame $B$, followed by the change from $B$ to $A$~\cite{Giacomini:2017zju,Ballesteros:2020lgl},
\begin{equation}\label{eq:ST_transitivity}
\hat S_T^{(C\to A)}=\hat S_T^{(B\to A)}\hat S_T^{(C\to B)}.
\end{equation}
It is useful to distinguish here two related algebraic structures. Ref.~\cite{Ballesteros:2020lgl} showed that the canonical transformations entering the dynamical QRF transformations close a group structure generated by a Lie algebra, whose generators provide the building blocks of transformations such as Eq.~\eqref{eq:STSb}. When the source and target reference frames are explicitly included in the description, the transformations $\hat S_T^{(C\to A)}$ can be naturally viewed as a groupoid.

To see this, let $\mathcal{F}$ denote the set of physical systems that can serve as QRFs, and associate to every pair $C,A\in\mathcal{F}$ the transformation $\hat S_T^{(C\to A)}$. Such transformations can be composed as changes of reference frame only when the target frame of the first transformation coincides with the source frame of the second. For the composable sequence $C\to B\to A$, Eq.~\eqref{eq:ST_transitivity} states that their composition is again the QRF transformation from $C$ to $A$. The identity transformation, amounting to a trivial change of perspective, is $\hat S_T^{(C\to C)}=\mathbb{1}$, while unitarity together with Eq.~\eqref{eq:ST_transitivity} implies
\begin{equation}
\hat S_T^{(A\to C)}=\big[\hat S_T^{(C\to A)}\big]^{-1}=\big[\hat S_T^{(C\to A)}\big]^\dagger.
\end{equation}
Associativity follows directly from associativity of operator composition. Therefore, the systems in $\mathcal{F}$ together with the invertible QRF transformations connecting their descriptions form a groupoid: in contrast to a group, the composition of two transformations is defined only when their source and target frames match. The same reasoning applies to the purely kinematical transformation $\hat S_x^{(C\to A)}$ of Eq.~\eqref{eq:Sx_def}, since it corresponds to $\hat S_T^{(C\to A)}$ for $t=0$.
Therefore, the position-QRF transformations inherit the same groupoid structure.

The groupoid structure of the frame-labelled transformations should thus be distinguished from the group structure identified in Ref.~\cite{Ballesteros:2020lgl} for the canonical transformations from which the QRF maps are constructed. The former keeps track of the change between different QRF descriptions, while the latter characterizes the underlying transformations acting on phase space.

Notice that this composition property concerns the QRF transformations themselves and is logically distinct from the prescriptions for transforming the dynamics, which will be discussed in Sec.~\ref{sec:extnonextDef}.
As discussed there, the inclusion of the term $i\hbar(\partial_t\hat S_T)\hat S_T^\dagger$ in the transformation of the Hamiltonian ensures that the resulting generator is compatible with successive time-dependent frame changes.

\subsection{Dynamical evolution}
\label{sec:extnonextDef}

The QRF transformation $\hat S_T^{(C\to A)}(t)$ introduced in Eq.~\eqref{eq:STSb} depends explicitly on time, since it accounts for the free evolution of the frames $A$ and $C$ themselves. This introduces a feature that is absent for the fixed-time transformation $\hat S_x^{(C\to A)}$ of Eq.~\eqref{eq:Sx_def}: a time-dependent change of QRF modifies the generator of time evolution not only through the usual unitary conjugation of the Hamiltonian, but also through the explicit time dependence of the transformation itself. One may therefore either retain the full time-dependent map $\hat S_T(t)$, which can realize an extended symmetry in the sense of Refs.~\cite{Giacomini:2017zju,Ballesteros:2020lgl}, or freeze the map at an initial time, $\hat S_T(0)=\hat S_x$, and subsequently use it as a time-independent change of variables. We call the latter the time-independent, initial-slice, construction.

To understand these two logically distinct approaches, let $\hat H^{(C)}_{AB}$ be the Hamiltonian generating time evolution in the frame $C$, so that a state $|\psi^{(C)}(t)\rangle$ satisfies the Schr\"odinger equation
\begin{equation}
i\hbar\, \partial_t |\psi^{(C)}(t)\rangle = \hat H^{(C)}_{AB} |\psi^{(C)}(t)\rangle .
\label{eq:Cframe-Schrodinger}
\end{equation}
Consider a time-dependent QRF transformation $\hat S_T(t)\equiv \hat S_T^{(C\to A)}(t)$, unitary for every $t$, and define the state in the new frame as
\begin{equation}
|\psi^{(A)}(t)\rangle := \hat S_T(t)\, |\psi^{(C)}(t)\rangle .
\label{eq:state-transform}
\end{equation}
\vspace{3mm}
\paragraph{Time-dependent (extended-symmetry) transformation.}
Differentiating Eq.~\eqref{eq:state-transform} and using Eq.~\eqref{eq:Cframe-Schrodinger} gives
\begin{align}
i\hbar\,\partial_t|\psi^{(A)}(t)\rangle
&=i\hbar\bigl(\partial_t\hat S_T\bigr)|\psi^{(C)}(t)\rangle+\hat S_T\,i\hbar\,\partial_t|\psi^{(C)}(t)\rangle \notag\\
&=\left[i\hbar\bigl(\partial_t\hat S_T\bigr)\hat S_T^\dagger+\hat S_T\hat H_{AB}^{(C)}\hat S_T^\dagger\right]|\psi^{(A)}(t)\rangle .
\label{eq:derivation-extended}
\end{align}
Hence $|\psi^{(A)}(t)\rangle$ obeys a Schr\"odinger equation with generator
\begin{equation}
\hat H_{BC}^{(A)}(t)=\hat S_T(t)\hat H_{AB}^{(C)}\hat S_T^\dagger(t)+i\hbar\bigl(\partial_t\hat S_T(t)\bigr)\hat S_T^\dagger(t).
\label{eq:extended-H}
\end{equation}
Equation~\eqref{eq:extended-H} is the general transformation law for the Hamiltonian under a time-dependent unitary change of QRF. Following Refs.~\cite{Giacomini:2017zju,Ballesteros:2020lgl}, such a QRF transformation realizes an \emph{extended symmetry} when the transformed Hamiltonian has the same functional form as the original one, with the labels of the old and new reference frames exchanged. In particular, for the free-particle Hamiltonian $\hat H_{AB}^{(C)}=(\hat p_A^{(C)})^2/(2m_A)+(\hat p_B^{(C)})^2/(2m_B)$, the transformation $\hat S_T^{(C\to A)}(t)$ is an extended symmetry because Eq.~\eqref{eq:extended-H} gives
\begin{equation}
\hat H_{BC}^{(A)}=\frac{\bigl(\hat p_B^{(A)}\bigr)^2}{2m_B}+\frac{\bigl(\hat p_C^{(A)}\bigr)^2}{2m_C}.
\end{equation}
Thus the full time dependence of the QRF map is taken into account, and the transformed generator reproduces the free dynamics in the new frame rather than only the instantaneous change of variables at a fixed time.

A time-dependent QRF transformation also induces explicit time dependence in transformed observables. Given a Schr\"odinger-picture operator $\hat O^{(C)}$ with no explicit time dependence in frame $C$, define its $A$-frame image as
\begin{equation}
\hat O_{S}^{(A)}(t):=\hat S_T(t)\hat O^{(C)}\hat S_T^\dagger(t).
\end{equation}
Its Heisenberg-picture representative, evolved with $\hat H_{BC}^{(A)}(t)$, therefore satisfies
\begin{equation}\label{eq:extended-Heisenberg}
\frac{d}{dt}\hat O_{H}^{(A)}(t)=\frac{i}{\hbar}\bigl[\hat H_{H}^{(A)}(t),\hat O_{H}^{(A)}(t)\bigr]+\left(\frac{\partial\hat O_{S}^{(A)}(t)}{\partial t}\right)_{H}.
\end{equation}
The second term accounts for the explicit time dependence inherited from the QRF transformation itself. This distinction will be important below: in the extended-symmetry construction the transformed Hamiltonian retains the free-particle form, while the relational observable of interest acquires an explicit time dependence through $\hat S_T(t)$.

The transformation law~\eqref{eq:extended-H} is also consistent under successive time-dependent changes of QRF. Indeed, let
\begin{equation}
\hat S_T^{(C\to A)}(t)=\hat S_T^{(B\to A)}(t)\hat S_T^{(C\to B)}(t).
\end{equation}
After the first transformation, $C\to B$, the Hamiltonian is
\begin{equation}
\hat H_{AC}^{(B)}=\hat S_T^{(C\to B)}\hat H_{AB}^{(C)}\hat S_T^{(C\to B)\dagger}+i\hbar\bigl(\partial_t\hat S_T^{(C\to B)}\bigr)\hat S_T^{(C\to B)\dagger}.
\end{equation}
Applying the same transformation law to the subsequent change $B\to A$ gives
\begin{align}
\hat H_{BC}^{(A)}
&=\hat S_T^{(B\to A)}\hat H_{AC}^{(B)}\hat S_T^{(B\to A)\dagger}
+i\hbar\bigl(\partial_t\hat S_T^{(B\to A)}\bigr)\hat S_T^{(B\to A)\dagger} \notag\\
&=\hat S_T^{(C\to A)}\hat H_{AB}^{(C)}\hat S_T^{(C\to A)\dagger}
+i\hbar\bigl(\partial_t\hat S_T^{(C\to A)}\bigr)\hat S_T^{(C\to A)\dagger}.
\end{align}
Thus the inhomogeneous term $i\hbar(\partial_t\hat S_T)\hat S_T^\dagger$ is precisely what makes the Hamiltonian transformation law compatible with the composition of time-dependent QRF changes. This compositional property concerns the transformation of the dynamics and should be distinguished from the algebraic group structure of the canonical transformations entering the QRF maps discussed in Ref.~\cite{Ballesteros:2020lgl}.

\vspace{3mm}
\paragraph{Time-independent (initial slice)  transformation.}

An alternative,  construction consists in using the transformation evaluated at a fixed instant, say $t=0$, to relate the two frames' operators, and treating the resulting generator as if it were time-independent. That is, one defines
\begin{equation}
\hat H_{BC}^{(A|C)} := \hat S_T(0)\, \hat H^{(C)}_{AB}\, \hat S_T^\dagger(0) ,
\label{eq:nonextended-H}
\end{equation}
with $\hat S_T(0)=\hat S_x$, and use this operator to evolve the transformed initial $A$-frame operators via the ordinary Heisenberg equation $\frac{d\hat O^{(A)}}{dt} = \frac{i}{\hbar}[\hat H_{BC}^{(A|C)}, \hat O^{(A)}(t)]$. We refer to this construction as a time-independent (initial slice)  transformation. 
Crucially, no term proportional to $\partial_t\hat S_T$ is included because the map is, by construction, frozen at $t=0$. Notice that this does not mean that $\partial_t\hat S_T(t)|_{t=0}$ vanishes. In fact, the Hamiltonian obtained from the full time-dependent transformation satisfies
\begin{equation}
\hat H^{(A)}_{BC}(0)=\hat H_{BC}^{(A|C)}+i\hbar\left.\big(\partial_t\hat S_T(t)\big)\hat S_T^\dagger(t)\right|_{t=0}.
\end{equation}
Thus $\hat H_{BC}^{(A|C)}$ is not the Hamiltonian produced by the extended-symmetry prescription, even at the initial time.
The initial-slice construction therefore retains only the
instantaneous change of variables at $t=0$, whereas the extended-symmetry construction accounts for the subsequent motion of the frame.
Because the explicit time dependence of the QRF map is discarded, this construction need not define an extended symmetry of the dynamics: in particular, $\hat H_{BC}^{(A|C)}$ need not retain the free-particle functional form of the original Hamiltonian, as will be seen explicitly in Eq.~\eqref{eq:H_tilde}.

Both constructions are nonetheless physically legitimate and, as we show explicitly in Sec.~\ref{sec:derivation-eq35-ST}, they reproduce the same relational trajectory, Eqs.~\eqref{eq:xB_A_eq35} and \eqref{eq:eomExtST_eq35}. In the initial-slice construction, this trajectory is represented by the Heisenberg evolution of $\hat x_B^{(A)}$ under the transformed Hamiltonian $\hat H_{BC}^{(A|C)}$, which contains an explicit cross-coupling between $B$ and $C$, Eq.~\eqref{eq:H_tilde_negative_mass_commuting}. In the extended-symmetry construction, instead, the bare position $\hat x_B^{(A)}$ evolves freely, while the same relational trajectory is represented by the explicitly time-dependent transformed observable $\hat{\mathcal Q}_{B|A}^{(A)}(t)$. The distinction between the two constructions is therefore a matter of how the same time dependence is distributed between the transformed observable and the generator of evolution, rather than a disagreement about relational predictions.

%%%%%%%%%%%%%%%%%%%%%%%%%%%%%%%%%%%%%%%%%%%%%%%%%%%%%%%%%%%%%%%%%%%%%
\section{QRF interpretation of trajectory sharpness}

The generalized Galilean transformations described in the previous Section can be used to identify the EPR observables with observables that, in the frame of the negative-mass particle, are local to distinct subsystems. We show this first via the instantaneous translation $\hat S_x$, which makes the correspondence transparent, and then extend the construction to the time-dependent transformation $\hat S_T$. This allows us to establish the correspondence at the dynamical level, relevant for the construction of sharp trajectories.

\subsection{Fixed-time QRF transformation $\hat S_x$}
\label{sec:intSxTransf}

As in Section~\ref{sec:toymod}, we consider two particles, $A$ and $B$, with $A$ the negative-mass particle and $B$ the probe particle. The relevant EPR observables, defined in the laboratory frame $C$ are
\begin{equation}\label{eq:EPRobservables}
 \hat  Q_{B|A}^{(C)}\equiv \xbc - \xac\,, \qquad\qquad
\hat \Pi_{B|A}^{(C)}\equiv\pbc + \pac .
\end{equation}

To make the QRF correspondence explicit, we apply to them the instantaneous QRF transformation $\hat S_x$. Using Eqs.~\eqref{eqs:SxDef} we find that the relative position and total momentum are mapped to operators local in the frame of $A$:
\begin{equation}\label{eq:SxEPRpair}
\begin{aligned}
 \hat Q_{B|A}^{(A)}&\equiv  \hat S_x \, \hat Q_{B|A}^{(C)} \, \hat S_x^\dagger  = \hat S_x \, (\xbc - \xac) \, \hat S_x^\dagger = \xba, \\
\hat \Pi_{B|A}^{(A)}&\equiv \hat S_x \,  \hat \Pi_{B|A}^{(C)} \, \hat S_x^\dagger=\hat S_x \, (\pbc + \pac) \, \hat S_x^\dagger = -\pca .
\end{aligned}
\end{equation}
Hence, a sharp relative coordinate $\xbc-\xac$, as described in frame $C$, is mapped to a sharp position $\xba$ of system $B$ in frame $A$. Similarly, a sharp total momentum $\pbc+\pac$ is mapped to a sharp momentum $\pca$ of system $C$ in frame $A$. Since these two transformed observables act on distinct subsystems, $[\xba,\pca]=0$, consistently with the fact that the original EPR pair commutes.
Therefore, the trajectory-sharpening effect can be interpreted as the existence of a quantum reference frame in which the uncertainties of the two commuting observables that define the trajectory can be made simultaneously arbitrarily small.

So far, this is a purely kinematical correspondence. Equation~\eqref{eq:SxEPRpair} relates the descriptions in the frames $C$ and $A$ at a fixed time, so it does not determine how the corresponding observables evolve. 
To recover the dynamical trajectory discussed in Section~\ref{sec:toymod}, we therefore evolve the relational trajectory in frame $A$, using the map of the free-particle Hamiltonian in frame $C$ to frame $A$, following the procedure outlined in Section \ref{sec:extnonextDef}.

The free Hamiltonian of $A$ and $B$ in frame $C$ is
\begin{equation}
\hat H^{(C)}_{AB} = \frac{(\pac)^2}{2 m_A} + \frac{(\pbc)^2}{2 m_B}\,.
\label{eq:H_signed_C}
\end{equation}

The action of the instantaneous transformation $\hat S_x$ on the relevant momenta is given by Eqs.~\eqref{eqs:SxDef} as
\begin{subequations}\label{eq:stp_zero}
\begin{align}
    \hat S_x
    \hat p_B^{(C)}
    \hat S_x^{\dagger}
    &=
    \hat p_B^{(A)},
    \label{eq:pB_map_ST0}
    \\
    \hat S_x \hat p_A^{(C)} \hat S_x^{\dagger}
    &= -\hat p_C^{(A)} - \hat p_B^{(A)}.
    \label{eq:pA_map_ST0}
\end{align}
\end{subequations}
Using these identities, the $C$-frame Hamiltonian written in $A$-frame variables is then
\begin{align}
    \hat H_{BC}^{(A|C)}
    & \equiv
    \hat S_x
    \hat H_{AB}^{(C)}
    \hat S_x^{\dagger}
    =
    \frac{\big(-\hat p_C^{(A)}-\hat p_B^{(A)}\big)^2}{2m_A}
    +
    \frac{\big(\hat p_B^{(A)}\big)^2}{2m_B}.
    \label{eq:H_tilde}
\end{align}
The operator $\hat H_{BC}^{(A|C)}$ should not be confused with the Hamiltonian obtained from the full time-dependent QRF transformation (Eq.~\eqref{eq:extHtr} in the following Subsection). Rather, it is the $C$-frame generator rewritten in $A$-frame variables using the unitary $\hat S_x$. Thus it generates, in the transformed variables, dynamics that is unitarily equivalent to the original $C$-frame evolution, but it need not retain the free-particle form of $\hat H_{AB}^{(C)}$. The notation $(A|C)$ is used to emphasize this distinction from the extended-symmetry Hamiltonian $\hat H_{BC}^{(A)}$ introduced below.

For $m_B=m>0$ and $m_A=-m$, the two terms in Eq.~\eqref{eq:H_tilde} proportional to $(\hat p_B^{(A)})^2$ cancel, and since $\hat p_B^{(A)}$ and $\hat p_C^{(A)}$ commute we obtain
\begin{equation}
\hat H_{BC}^{(A|C)}= - \frac{\big(\hat p_C^{(A)}\big)^2}{2m} - \frac{1}{m}\hat p_C^{(A)}\hat p_B^{(A)}.
\label{eq:H_tilde_negative_mass_commuting}
\end{equation}
In the remainder of this Subsection, time-dependent operators are understood in the Heisenberg picture generated by $\hat H_{BC}^{(A|C)}$. In this picture, the Hamiltonian of Eq.~\eqref{eq:H_tilde_negative_mass_commuting} generates the time evolution of the relative coordinate $\hat Q_{B|A}^{(A)}=\hat x_B^{(A)}(0)$.
Using the commutation relation $\big[\hat x_B^{(A)},\hat p_B^{(A)}\big]=i\hbar$ in frame $A$, we get
\begin{align}
    \frac{d}{dt}\hat x_B^{(A)}(t)
    &= \frac{i}{\hbar} \left[ \hat H_{BC}^{(A|C)}, \hat x_B^{(A)}(t) \right] \nonumber\\
    &= \frac{i}{\hbar} \left[ -\frac{1}{m}\hat p_C^{(A)}\hat p_B^{(A)}, \hat x_B^{(A)}(t) \right] \nonumber\\
    &= -\frac{1}{m}\hat p_C^{(A)}(t). \label{eq:xB_A_velocity}
\end{align}
The momentum \(\hat p_C^{(A)}\) is conserved under \(\hat H_{BC}^{(A|C)}\), because the Hamiltonian depends only on commuting momenta.
Thus $\hat p_C^{(A)}(t)=\hat p_C^{(A)}(0)$ and integrating Eq.~\eqref{eq:xB_A_velocity} we obtain
\begin{equation}
 \hat x_B^{(A)}(t) = \hat x_B^{(A)}(0) - \frac{t}{m}\hat p_C^{(A)}(0) .  \label{eq:xB_A_eq35}
\end{equation}

Interestingly, the dynamics of the position of $B$ in the frame of the negative-mass system $A$ is governed by the momentum $\hat p_C^{(A)}$ of the distinct subsystem $C$, rather than by the canonical momentum $\hat p_B^{(A)}$ conjugate to $\hat x_B^{(A)}$. The origin of this feature is transparent in this derivation.
In frame $C$, the velocity of the relative coordinate $\hat Q_{B|A}^{(C)}$ is proportional to the total momentum $\hat\Pi_{B|A}^{(C)}=\hat p_B^{(C)}+\hat p_A^{(C)}$ when $m_A=-m_B$, see Eq.~\eqref{eq:EOMrelX}.
Under the instantaneous transformation $\hat S_x$, the total momentum is mapped according to Eqs.~\eqref{eq:SxEPRpair}: $\hat p_B^{(C)}+\hat p_A^{(C)} \mapsto -\hat p_C^{(A)}$.
This is consistent with the presence in the transformed Hamiltonian \(\hat H_{BC}^{(A|C)}\) of the cross term \(-\hat p_C^{(A)}\hat p_B^{(A)}/m\), which makes \(\hat x_B^{(A)}\) evolve with velocity \(-\hat p_C^{(A)}/m\).

\subsection{Time-dependent QRF transformation $\hat S_T$}
\label{sec:derivation-eq35-ST}

The instantaneous and dynamical equivalence between the EPR observables and observables local to the negative-mass particle can be demonstrated also through the generalized Galilean translation $\hat S_T^{(C\to A)}$ from the $C$ frame to the negative-mass frame $A$. 
With respect to the construction of the previous Subsection, in this case the time dependence is distributed differently: while in the $\hat S_x$ mapping the time evolution of the relational observable is fully determined by the transformed Hamiltonian, in the $\hat S_T$ construction of this Subsection the time evolution is partly determined by the time-dependent transformations and partly by the Hamiltonian. Nevertheless, the two constructions describe the same relational trajectory.

In contrast to the derivation in the previous Subsection, where we first transformed the initial relational coordinate to frame $A$ using $\hat S_x^{(C\to A)} \equiv \hat S_T^{(C\to A)}(0)$, and then evolved the transformed initial operator with the corresponding transformed Hamiltonian
\begin{equation}
\hat H_{BC}^{(A|C)}=\hat S_T^{(C\to A)}(0)\hat H_{AB}^{(C)}\hat S_T^{(C\to A)\dagger}(0),
\end{equation}
here we adopt the full time-dependent transformation $\hat S_T(t)$, which is an extended symmetry of the free-particle Hamiltonian of Eq.~\eqref{eq:H_signed_C}. Therefore, in this case, the transformed Hamiltonian retains the free-particle form in the new frame,
\begin{equation} \label{eq:extHtr}
\begin{aligned}
\hat H_{BC}^{(A)} &= \hat S_T^{(C\to A)}(t)\hat H_{AB}^{(C)} \hat S_T^{(C\to A)\dagger}(t) +i\hbar\left(\partial_t\hat S_T^{(C\to A)}(t)\right) \hat S_T^{(C\to A)\dagger}(t) \\
&= \frac{\bigl(\hat p_B^{(A)}\bigr)^2}{2m_B} + \frac{\bigl(\hat p_C^{(A)}\bigr)^2}{2m_C},
\end{aligned}
\end{equation}
and the cross term present in $\hat H_{BC}^{(A|C)}$ is absent.

The time-dependent image, under the full time-dependent QRF transformation, of the initial relational coordinate $\hat Q_{B|A}^{(C)}(0)$ is
\begin{equation}
\hat{\mathcal Q}_{B|A,S}^{(A)}(t) := \hat S_T^{(C\to A)}(t)\, \hat Q_{B|A}^{(C)}(0)\, \hat S_T^{(C\to A)\dagger}(t).
\label{eq:QrelA_def}
\end{equation}
Using Eqs.~\eqref{eq:DefSTtransf} and $m_A=-m_B=-m$, one finds
\begin{equation}\label{eq:QrelA_explicit}
\hat{\mathcal Q}_{B|A,S}^{(A)}(t) = \hat x_B^{(A)} - \frac{t}{m} \left(\hat p_B^{(A)}+\hat p_C^{(A)}\right).
\end{equation}
We used the additional subscript $S$ to emphasize that at this stage, $\hat{\mathcal Q}_{B|A}^{(A)}(t)$ denotes the transformed image of an observable in the Schr\"odinger representation, whose explicit time dependence is inherited from the time-dependent QRF transformation. 
The Heisenberg-picture evolution of $\hat{\mathcal Q}_{B|A}^{(A)}$ is generated by the transformed Hamiltonian, Eq.~\eqref{eq:extHtr}, according to the time-evolution equation~\eqref{eq:extended-Heisenberg}: 
\begin{equation}
\frac{d}{dt}\hat{\mathcal Q}_{B|A,H}^{(A)}(t)=\frac{i}{\hbar}\left[\hat H_{BC}^{(A)},\hat{\mathcal Q}_{B|A,H}^{(A)}(t)\right]+\left(\frac{\partial\hat{\mathcal Q}_{B|A,S}^{(A)}(t)}{\partial t}\right)_H= -\frac{1}{m}\hat p_C^{(A)}(0)\,,
\end{equation}
where in the last equality we used 
 Eqs.~\eqref{eq:extHtr} and \eqref{eq:QrelA_explicit} to write the two contributions as
\begin{equation}
\frac{i}{\hbar}\left[\hat H_{BC}^{(A)},\hat{\mathcal Q}_{B|A,H}^{(A)}(t)\right]=\frac{1}{m}\hat p_B^{(A)}(0),\qquad \left(\frac{\partial\hat{\mathcal Q}_{B|A}^{(A)}(t)}{\partial t}\right)_H=-\frac{1}{m}\left(\hat p_B^{(A)}(0)+\hat p_C^{(A)}(0)\right).
\end{equation}
Since $\hat{\mathcal Q}_{B|A}^{(A)}(0)=\hat x_B^{(A)}(0)$, see Eq.~\eqref{eq:SxEPRpair}, and both momenta are conserved under $\hat H_{BC}^{(A)}$, integration gives
\begin{equation}\label{eq:eomExtST_eq35}
\hat{\mathcal Q}_{B|A,H}^{(A)}(t)=\hat x_B^{(A)}(0)-\frac{t}{m}\hat p_C^{(A)}(0).
\end{equation}

Equation~\eqref{eq:eomExtST_eq35} coincides with the relational trajectory obtained in the initial-slice construction, Eq.~\eqref{eq:xB_A_eq35}. The agreement concerns the same relational observable, not the free evolution of $\hat x_B^{(A)}$ under $\hat H_{BC}^{(A)}$. In the initial-slice construction, the relational trajectory is represented by the Heisenberg evolution of $\hat x_B^{(A)}$ under the non-free Hamiltonian $\hat H_{BC}^{(A|C)}$, whereas in the extended-symmetry construction the same relational trajectory is represented by the explicitly time-dependent observable $\hat{\mathcal Q}_{B|A}^{(A)}(t)$, while the position operator $\hat x_B^{(A)}$, which coincides with $\hat{\mathcal Q}_{B|A}^{(A)}$ only at $t=0$, evolves freely. Indeed, for $m_B=m$, under the free Hamiltonian~\eqref{eq:extHtr} the latter evolves as
\begin{equation}
\hat x_{B,H}^{(A)}(t)=\hat x_B^{(A)}(0)+\frac{t}{m}\hat p_B^{(A)}(0).
\end{equation}

\section{QRF dependence of trajectory sharpness}

Having established the QRF interpretation of EPR variables, in this Section we investigate how the trajectory-sharpening effect depends on the choice of QRF when multiple candidate reference systems are available.

As a first step, we define two QRF-invariant measures of relational trajectory sharpness: an instantaneous marginal uncertainty product and a covariance-area measure that also accounts for correlations between the relevant observables. We show that, under opposite-mass free evolution, the covariance area is conserved and equals the minimum value attained by the marginal uncertainty product along the trajectory. We then derive a monogamy relation showing that a probe trajectory cannot be made arbitrarily sharp simultaneously with respect to two distinct negative-mass reference systems, even when each trajectory is considered at its own optimal focus time. Finally, we show that this restriction does not preclude sharpness from being encoded in a collective reference mode constructed from two physical systems, extending the trajectory-sharpening construction beyond individual reference systems.

\subsection{Trajectory sharpness measure and QRF invariance}
\label{sec:sharpnessDef}

\begin{figure}[t!]
    \centering
    \includegraphics[width=0.9\linewidth]{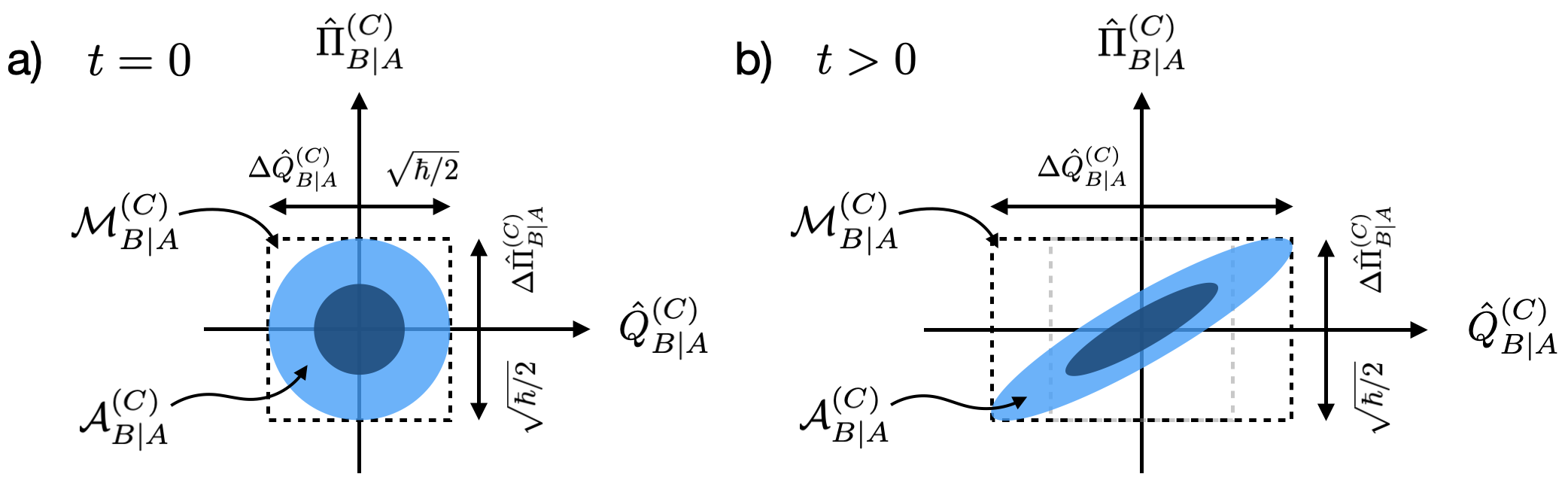}
    \caption{\textbf{Relational trajectory sharpness.} Snapshots of the joint distribution of the commuting relational observables $\hat Q_{B|A}^{(C)}$ and $\hat\Pi_{B|A}^{(C)}$ at two different times under opposite-mass free evolution. The marginal product $\mathcal M_{B|A}^{(C)}(t)$ of Eq.~\eqref{eq:M_BA_def} is proportional to the area of the axis-aligned uncertainty rectangle determined by the two marginal standard deviations, while $\mathcal A_{B|A}^{(C)}(t)$ of Eq.~\eqref{eq:AreaDef} is proportional to the area of the corresponding covariance ellipse. Both quantities are invariant under QRF transformations. Under free opposite-mass evolution, $\mathcal A_{B|A}$ is exactly conserved, whereas $\mathcal M_{B|A}(t)$ changes whenever the covariance between the two observables changes. Since $[\hat Q_{B|A}^{(C)},\hat\Pi_{B|A}^{(C)}]=0$, there is no nonzero uncertainty-relation lower bound on either quantity. The value $\hbar/2$ shown in panel (a) is only a reference scale and is not a lower bound.}
    \label{fig:2}
\end{figure}

The relational trajectory of a free system $B$ with respect to another free system $A$ with opposite mass, as described in the laboratory frame $C$, is determined by a commuting pair of EPR observables $(\hat Q_{B|A}^{(C)},\hat\Pi_{B|A}^{(C)})$ defined in Eq.~\eqref{eq:EPRobservables},
\begin{equation}
[\hat Q_{B|A}^{(C)}(t),\hat\Pi_{B|A}^{(C)}(t)]=0\,,
\end{equation}
and can therefore be made arbitrarily sharp, as there is no uncertainty-relation lower bound preventing their uncertainties from being made simultaneously arbitrarily small. Their marginal uncertainty area is quantified by
\begin{equation}\label{eq:M_BA_def}
\mathcal M_{B|A}^{(C)}(t):=\Delta\hat Q_{B|A}^{(C)}(t)\,\Delta\hat\Pi_{B|A}^{(C)}(t)\geq0,
\end{equation}
where $\Delta\hat O$ denotes the standard deviation of the observable $\hat O$. Geometrically, $\mathcal M_{B|A}^{(C)}$ is proportional to the area of the axis-aligned uncertainty rectangle in the $(Q,\Pi)$ plane, see Fig.~\ref{fig:2}.
Notice that $\mathcal M_{B|A}^{(C)}\to0$ alone does not imply that both marginal uncertainties vanish separately, therefore trajectory sharpness, such that $\Delta\hat Q_{B|A}^{(C)}\to0$ and $\Delta\hat\Pi_{B|A}^{(C)}\to0$ simultaneously, is a stronger condition.

The uncertainty area $\mathcal M_{B|A}$ is invariant under QRF change of description at any fixed time. To see this, consider first the three-system setting consisting only of $B$, $A$, and the laboratory frame $C$, and let $\hat S_x^{(C\to A)}$ be the instantaneous QRF transformation from frame $C$ to frame $A$. We work in the Heisenberg picture, in which states are time independent and observables carry the full time dependence. In frame $C$, $\hat O^{(C)}(t)=\hat U_C^\dagger(t)\,\hat O^{(C)}\,\hat U_C(t)$ with $\hat U_C(t)=e^{-\frac{i}{\hbar}\hat H_{AB}^{(C)}t}$, while in frame $A$ observables evolve under the initial-slice Hamiltonian $\hat H_{BC}^{(A|C)}=\hat S_x^{(C\to A)}\hat H_{AB}^{(C)}\hat S_x^{(C\to A)\dagger}$ of Eq.~\eqref{eq:H_tilde}. Since $\hat S_x^{(C\to A)}$ is time independent, $e^{-\frac{i}{\hbar}\hat H_{BC}^{(A|C)}t}=\hat S_x^{(C\to A)}\hat U_C(t)\hat S_x^{(C\to A)\dagger}$, and therefore at any fixed time $t$ the state and observables transform as
\begin{equation}
\hat\rho^{(A)}=\hat S_x^{(C\to A)}\hat\rho^{(C)}\hat S_x^{(C\to A)\dagger},\qquad \hat O^{(A)}(t)=\hat S_x^{(C\to A)}\hat O^{(C)}(t)\hat S_x^{(C\to A)\dagger}\,.
\end{equation}
Consequently, expectation values are preserved by the QRF transformation at any time,
\begin{equation}\label{eq:exp_val}
\mathrm{Tr}\!\left[\hat\rho^{(A)}\hat O^{(A)}(t)\right]=\mathrm{Tr}\!\left[\hat\rho^{(C)}\hat O^{(C)}(t)\right],
\end{equation}
and the same holds for all moments of the transformed observable, since $\hat S_x^{(C\to A)}\hat O^n\hat S_x^{(C\to A)\dagger}=\big(\hat S_x^{(C\to A)}\hat O\,\hat S_x^{(C\to A)\dagger}\big)^n$.
By the result above, the unitary $\hat S_x^{(C\to A)}$ maps the evolved relational observables in frame $C$ to their $A$-frame images at any fixed time $t$,
\begin{equation}
\hat Q_{B|A}^{(C)}(t)\mapsto\hat x_B^{(A)}(t),\qquad \hat\Pi_{B|A}^{(C)}(t)\mapsto-\hat p_C^{(A)}(t),
\end{equation}
where $\hat x_B^{(A)}(t)$ and $\hat p_C^{(A)}(t)$ denote the Heisenberg evolution under $\hat H^{(A|C)}_{BC}$ (initial-slice construction). Indeed, using Eq.~\eqref{eq:EOMrelX}, one finds $\hat S_x^{(C\to A)}\hat Q_{B|A}^{(C)}(t)\hat S_x^{(C\to A)\dagger}=\hat x_B^{(A)}(0)-\frac{t}{m}\hat p_C^{(A)}(0)$, which coincides with Eq.~\eqref{eq:xB_A_eq35}. In the extended-symmetry construction of Sec.~\ref{sec:derivation-eq35-ST}, the same image is represented by $\hat{\mathcal Q}_{B|A,H}^{(A)}(t)$ of Eq.~\eqref{eq:eomExtST_eq35}, and not by the freely evolving canonical position $\hat x_B^{(A)}(t)$.
Therefore, taking Eq.~\eqref{eq:exp_val} into account,
\begin{equation}
\Delta\hat Q_{B|A}^{(C)}(t)=\Delta\hat x_B^{(A)}(t),\qquad \Delta\hat\Pi_{B|A}^{(C)}(t)=\Delta\hat p_C^{(A)}(t),
\end{equation}
and it is natural to define
\begin{equation}
\mathcal M_{B|A}^{(A)}(t):=\Delta\hat x_B^{(A)}(t)\,\Delta\hat p_C^{(A)}(t).
\end{equation}
Hence
\begin{equation}\label{eq:MinvQRF}
\mathcal M_{B|A}^{(C)}(t)=\mathcal M_{B|A}^{(A)}(t).
\end{equation}
Thus we have shown that $\mathcal M_{B|A}$ is invariant under the QRF change of description at any fixed time, using an explicit $A$-frame representation that applies to the three-system setting. In a larger multipartite system, the same invariance holds when the complete QRF transformation is applied to the full relational pair, but the transformed momentum observable also contains the momenta of spectator subsystems. 
These contributions follow directly from the multipartite form of the position-QRF transformation. 
In general, if $\mathcal S$ denotes the set of non-frame spectator systems distinct from $B$, $C$, and $A$, then $\hat\Pi_{B|A}^{(C)}\longmapsto-\hat p_C^{(A)}-\sum_{R\in\mathcal S}\hat p_R^{(A)}$.

The marginal product $\mathcal M_{B|A}$ does not, however, retain information about correlations between the two commuting trajectory observables. To include these correlations, let us define
\begin{equation}\label{eq:cm_vectors}
\hat{\boldsymbol R}_{B|A}^{(C)}(t):=\begin{pmatrix}\hat Q_{B|A}^{(C)}(t)\\ \hat\Pi_{B|A}^{(C)}(t)\end{pmatrix},\qquad \hat{\boldsymbol R}_{B|A}^{(A)}(t):=\begin{pmatrix}\hat x_B^{(A)}(t)\\ -\hat p_C^{(A)}(t)\end{pmatrix}.
\end{equation}
For any operator vector $\hat{\boldsymbol R}$, the associated covariance matrix is
\begin{equation}\label{eq:cmDef}
\Gamma_{\mu\nu}(\hat{\boldsymbol R}):=\frac{1}{2}\left\langle\left\{\delta\hat R_\mu,\delta\hat R_\nu\right\}\right\rangle,\qquad \delta\hat R_\mu:=\hat R_\mu-\langle\hat R_\mu\rangle\,.
\end{equation}
Since $\hat S_x^{(C\to A)}$ is unitary, and both the operators and the states transform under it, i.e.\ $\hat{\boldsymbol R}_{B|A}^{(A)}(t)=\hat S_x^{(C\to A)}\hat{\boldsymbol R}_{B|A}^{(C)}(t)\hat S_x^{(C\to A)\dagger}$ and $\hat\rho^{(A)}=\hat S_x^{(C\to A)}\hat\rho^{(C)}\hat S_x^{(C\to A)\dagger}$, the expectation values of the fluctuations and of their anticommutators are preserved. Hence
\begin{equation}\label{eq:cmEq}
\Gamma_{B|A}^{(C)}(t)=\Gamma_{B|A}^{(A)}(t),
\end{equation}
where the two covariance matrices are evaluated in the corresponding QRF states.

The covariance matrix can be used to define a covariance-area sharpness parameter
\begin{equation}\label{eq:AreaDef}
\mathcal A_{B|A}^{(C)}(t):=\sqrt{\det\Gamma_{B|A}^{(C)}(t)}.
\end{equation}
This inherits invariance under instantaneous QRF transformations from the covariance matrix itself, Equation~\eqref{eq:cmDef}:
\begin{equation}
\mathcal A_{B|A}^{(C)}(t)=\mathcal A_{B|A}^{(A)}(t).
\end{equation}
Note that if one uses the pair $(\hat x_B^{(A)},\hat p_C^{(A)})$ instead of $(\hat x_B^{(A)},-\hat p_C^{(A)})$, the off-diagonal covariance changes sign, but both $\mathcal M_{B|A}$ and $\det\Gamma_{B|A}$ remain unchanged.

We now study the free evolution of these quantities. For $m_B=m>0$ and $m_A=-m$, Eq.~\eqref{eq:EOMrelX} generalizes to
\begin{equation}\label{eq:R_free_evolution}
\hat Q_{B|A}^{(C)}(t)=\hat Q_{B|A}^{(C)}(0)+\frac{t}{m}\hat\Pi_{B|A}^{(C)}(0),\qquad \hat\Pi_{B|A}^{(C)}(t)=\hat\Pi_{B|A}^{(C)}(0).
\end{equation}
It follows that
\begin{equation}\label{eq:VarQt}
\Delta^2\hat Q_{B|A}^{(C)}(t)=\Delta^2\hat Q_{B|A}^{(C)}(0)+\frac{2t}{m}\Cov\!\left[\hat Q_{B|A}^{(C)}(0),\hat\Pi_{B|A}^{(C)}(0)\right]+\frac{t^2}{m^2}\Delta^2\hat\Pi_{B|A}^{(C)}(0),
\end{equation}
while
\begin{equation}\label{eq:CovQt_general}
\Cov\!\left[\hat Q_{B|A}^{(C)}(t),\hat\Pi_{B|A}^{(C)}(t)\right]=\Cov\!\left[\hat Q_{B|A}^{(C)}(0),\hat\Pi_{B|A}^{(C)}(0)\right]+\frac{t}{m}\Delta^2\hat\Pi_{B|A}^{(C)}(0).
\end{equation}
Thus the marginal width need not increase monotonically: depending on the initial covariance, the relational position distribution may initially contract before subsequently expanding.

The free evolution~\eqref{eq:R_free_evolution} is a shear,
\begin{equation}
\hat{\boldsymbol R}_{B|A}^{(C)}(t)=M(t)\hat{\boldsymbol R}_{B|A}^{(C)}(0),\qquad M(t):=\begin{pmatrix}1&t/m\\0&1\end{pmatrix},
\end{equation}
and therefore $\Gamma_{B|A}^{(C)}(t)=M(t)\Gamma_{B|A}^{(C)}(0)M(t)^T$.
Since $\det M(t)=1$, the determinant of the covariance matrix is exactly conserved, giving
\begin{equation}\label{eq:Atinv}
\mathcal A_{B|A}^{(C)}(t)=\mathcal A_{B|A}^{(C)}(0)\qquad\text{for all }t.
\end{equation}
By QRF invariance, the same constant value is obtained in the corresponding $A$-frame description.

The relation between the two sharpness parameters is particularly simple. By definition,
\begin{equation}\label{eq:M_A_relation}
\big[\mathcal M_{B|A}^{(C)}(t)\big]^2=\big[\mathcal A_{B|A}^{(C)}(t)\big]^2+\Cov\!\left[\hat Q_{B|A}^{(C)}(t),\hat\Pi_{B|A}^{(C)}(t)\right]^2.
\end{equation}
Hence,
\begin{equation}
\mathcal A_{B|A}^{(C)}(t)\leq\mathcal M_{B|A}^{(C)}(t),
\end{equation}
with equality if and only if the two relational observables are uncorrelated at that time. Equation~\eqref{eq:M_A_relation} makes explicit that the time dependence of $\mathcal M_{B|A}$ is entirely due to the changing covariance between the two trajectory observables, while $\mathcal A_{B|A}$ removes this shear-dependent contribution and remains constant: the free evolution shears the covariance ellipse, changing its bounding rectangle but not its area, as illustrated in Fig.~\ref{fig:2}.

For $\Delta^2\hat\Pi_{B|A}^{(C)}(0)>0$, Eq.~\eqref{eq:CovQt_general} shows that there is a unique time
\begin{equation}\label{eq:t_focus}
t_*=-m\,\frac{\Cov\!\left[\hat Q_{B|A}^{(C)}(0),\hat\Pi_{B|A}^{(C)}(0)\right]}{\Delta^2\hat\Pi_{B|A}^{(C)}(0)}
\end{equation}
at which the covariance vanishes. At this time the relational position variance is minimal, $\Delta^2\hat Q_{B|A}^{(C)}(t_*) = \big[\mathcal A_{B|A}^{(C)}\big]^2 / \Delta^2\hat\Pi_{B|A}^{(C)}$,
and Eq.~\eqref{eq:M_A_relation} gives $\mathcal M_{B|A}^{(C)}(t_*)=\mathcal A_{B|A}^{(C)}$. 
Therefore, over the full free evolution, $\mathcal A_{B|A}^{(C)}=\min_{t\in\mathbb R}\mathcal M_{B|A}^{(C)}(t)$.
The special case of vanishing covariance at $t=0$ corresponds to the special case in which the chosen initial slice coincides with this minimum-width, or focus, time. In that case,
\begin{equation}
\Delta^2\hat Q_{B|A}^{(C)}(t)=\Delta^2\hat Q_{B|A}^{(C)}(0)+\frac{t^2}{m^2}\Delta^2\hat\Pi_{B|A}^{(C)}(0),
\end{equation}
so $\mathcal M_{B|A}^{(C)}(t)\geq\mathcal M_{B|A}^{(C)}(0)=\mathcal A_{B|A}^{(C)}$ for all $t$. 
If, on the other hand, $\Delta^2\hat\Pi_{B|A}^{(C)}=0$, the Cauchy--Schwarz inequality $\big|\Cov\big[\hat Q_{B|A}^{(C)},\hat\Pi_{B|A}^{(C)}\big]\big|\le\Delta\hat Q_{B|A}^{(C)}\,\Delta\hat\Pi_{B|A}^{(C)}$ implies that the covariance vanishes at all times, consistently with Eq.~\eqref{eq:CovQt_general}. Hence $\mathcal M_{B|A}^{(C)}(t)=\mathcal A_{B|A}^{(C)}=0$, and Eq.~\eqref{eq:VarQt} shows that $\Delta\hat Q_{B|A}^{(C)}(t)=\Delta\hat Q_{B|A}^{(C)}(0)$ is constant, so there is no unique focus time.

The invariant area also has a direct trajectory interpretation. Since $\Delta\hat\Pi_{B|A}^{(C)}$ is conserved, the minimum positional width and the uncertainty in the relational velocity are
\begin{equation}
\Delta \hat Q_{B|A,\min}^{(C)}=\frac{\mathcal A_{B|A}^{(C)}}{\Delta\hat\Pi_{B|A}^{(C)}},\qquad \Delta \hat v_{B|A}^{(C)}=\frac{\Delta\hat\Pi_{B|A}^{(C)}}{m}.
\end{equation}
The covariance-area parameter \(\mathcal A_{B|A}\) is invariant under the instantaneous QRF transformation and conserved under the opposite-mass free evolution. Moreover, it equals the minimum value attained by the marginal uncertainty product \(\mathcal M_{B|A}(t)\) over the relational trajectory. It therefore provides a natural invariant measure of the covariance area associated with trajectory sharpness. The quantity $\mathcal M_{B|A}(t)$ remains useful as it quantifies the instantaneous marginal uncertainties, while $\mathcal A_{B|A}$ removes the dependence on the instantaneous correlation between the two trajectory observables.
Neither quantity alone guarantees that both marginal uncertainties are simultaneously small. This stronger condition of trajectory sharpness requires the separate suppression of \(\Delta \hat Q_{B|A}\) and \(\Delta\hat\Pi_{B|A}\).

%%%%%%%%%%%%%%%%%%%%%%%%%%%%%%%%%%%%%%%%%%%%%%%%%%%%%%%%%%%%%%%%%%%%%
\subsection{Monogamy relation for relational trajectory sharpness}

Having introduced QRF-invariant measures of relational trajectory sharpness, we now show that the trajectory of a given probe system $B$ cannot be made simultaneously arbitrarily sharp with respect to two distinct reference systems $A$ and $D$. The resulting trade-off follows directly from the algebra of the corresponding relational observables and holds for arbitrary states.

We consider four systems $A,B,C,D$, with $C$ serving as the initial reference frame. We take the probe system $B$ to have positive mass $m_B=m>0$, while the two candidate reference systems have opposite mass, $m_A=m_D=-m$. In frame $C$, the free Hamiltonian is
\begin{equation}
\hat H_{ABD}^{(C)}=\frac{\bigl(\hat p_A^{(C)}\bigr)^2}{2m_A}+\frac{\bigl(\hat p_B^{(C)}\bigr)^2}{2m_B}+\frac{\bigl(\hat p_D^{(C)}\bigr)^2}{2m_D}.
\end{equation}
For $F\in\{A,D\}$, Eq.~\eqref{eq:R_free_evolution} gives
\begin{equation}\label{eq:QBF_evolution_monogamy}
\hat Q_{B|F}^{(C)}(t)=\hat Q_{B|F}^{(C)}(0)+\frac{t}{m}\hat\Pi_{B|F}^{(C)}(0),\qquad \hat\Pi_{B|F}^{(C)}(t)=\hat\Pi_{B|F}^{(C)}(0).
\end{equation}
For either choice of $F$, the two observables defining the corresponding relational trajectory commute,
\begin{equation}
\bigl[\hat Q_{B|F}^{(C)}(t),\hat\Pi_{B|F}^{(C)}(t)\bigr]=0,
\end{equation}
and may therefore be made jointly arbitrarily sharp. The situation changes when the relational observables associated with the two different candidate frames are considered simultaneously. 

Although each relational pair consists of commuting observables, the crossed commutators show that the two pairs are not jointly compatible: the position relative to one candidate frame is canonically conjugate to the momentum-like observable associated with the other.
Indeed, the crossed commutators are
\begin{align}
\bigl[\hat Q_{B|A}^{(C)}(t),\hat\Pi_{B|D}^{(C)}(t)\bigr]&=i\hbar,\label{eq:crossed_commutator_1}\\
\bigl[\hat Q_{B|D}^{(C)}(t),\hat\Pi_{B|A}^{(C)}(t)\bigr]&=i\hbar.\label{eq:crossed_commutator_2}
\end{align}
The remaining commutators between the two relational pairs vanish,
\begin{equation}
\bigl[\hat Q_{B|A}^{(C)}(t),\hat Q_{B|D}^{(C)}(t)\bigr]=0,\qquad \bigl[\hat\Pi_{B|A}^{(C)}(t),\hat\Pi_{B|D}^{(C)}(t)\bigr]=0.
\end{equation}
Thus, although each trajectory pair is internally commuting, the two pairs are not mutually compatible.

Applying the Robertson uncertainty relation to the two crossed pairs gives
\begin{equation}\label{eq:crossed_Robertson}
\Delta\hat Q_{B|A}^{(C)}(t)\,\Delta\hat\Pi_{B|D}^{(C)}(t)\geq\frac{\hbar}{2},\qquad \Delta\hat Q_{B|D}^{(C)}(t)\,\Delta\hat\Pi_{B|A}^{(C)}(t)\geq\frac{\hbar}{2}.
\end{equation}
Multiplying the two inequalities and using Eq.~\eqref{eq:M_BA_def} yields
\begin{equation}\label{eq:M_monogamy_final}
\mathcal M_{B|A}^{(C)}(t)\,\mathcal M_{B|D}^{(C)}(t)\geq\frac{\hbar^2}{4}.
\end{equation}
This relation holds at every time $t$. In particular, as \(M_{B|A}(t)\to0\), the bound forces \(M_{B|D}(t)\) to diverge,
\begin{equation}
\mathcal M_{B|D}^{(C)}(t)\geq\frac{\hbar^2}{4\,\mathcal M_{B|A}^{(C)}(t)}\,.
\end{equation}
Thus the two relational trajectories may both have finite sharpness, but they cannot both be made arbitrarily sharp.

A stronger monogamy relation can be obtained for the covariance-area parameter $\mathcal A_{B|A}$ introduced in Eq.~\eqref{eq:AreaDef}. To obtain a bound that also accounts for correlations within each relational pair, we combine the two crossed canonical pairs into a single two-mode covariance inequality. To do this, define the two-component operator vectors
\begin{equation}\label{eq:XY_area_monogamy}
\hat{\boldsymbol X}(t):=\begin{pmatrix}\hat Q_{B|A}^{(C)}(t)\\ \hat\Pi_{B|A}^{(C)}(t)\end{pmatrix},\qquad \hat{\boldsymbol Y}(t):=\begin{pmatrix}\hat\Pi_{B|D}^{(C)}(t)\\ -\hat Q_{B|D}^{(C)}(t)\end{pmatrix}.
\end{equation}
Using Eqs.~\eqref{eq:crossed_commutator_1} and \eqref{eq:crossed_commutator_2}, together with the vanishing remaining commutators, one obtains
\begin{equation}\label{eq:vector_canonical}
[\hat X_i(t),\hat X_j(t)]=0,\qquad [\hat Y_i(t),\hat Y_j(t)]=0,\qquad [\hat X_i(t),\hat Y_j(t)]=i\hbar\,\delta_{ij}.
\end{equation}
Hence $\hat{\boldsymbol X}$ and $\hat{\boldsymbol Y}$ constitute a two-dimensional pair of canonically conjugate vector observables.

Let $\Gamma_X(t):=\mathrm{Cov}\!\left(\hat{\boldsymbol X}(t)\right)$ and $\Gamma_Y(t):=\mathrm{Cov}\!\left(\hat{\boldsymbol Y}(t)\right)$.
For arbitrary real vectors $\boldsymbol u,\boldsymbol v \in\mathbb R^2$, the observables $\hat X_{\boldsymbol u}:=\boldsymbol u^T\hat{\boldsymbol X}$, $\hat Y_{\boldsymbol v}:=\boldsymbol v^T\hat{\boldsymbol Y}$, satisfy $[\hat X_{\boldsymbol u},\hat Y_{\boldsymbol v}]=i\hbar\,\boldsymbol u^T\boldsymbol v$.
Robertson's uncertainty relation therefore gives
\begin{equation}\label{eq:vector_Robertson}
\bigl(\boldsymbol u^T\Gamma_X\boldsymbol u\bigr)\bigl(\boldsymbol v^T\Gamma_Y\boldsymbol v\bigr)\geq\frac{\hbar^2}{4}\bigl(\boldsymbol u^T\boldsymbol v\bigr)^2
\end{equation}
for all real $\boldsymbol u,\boldsymbol v$.
For fixed $\boldsymbol v$, Eq.~\eqref{eq:vector_Robertson} implies
\begin{equation}
\boldsymbol v^T\Gamma_Y\boldsymbol v\geq\frac{\hbar^2}{4}\sup_{\boldsymbol u\neq0}\frac{\bigl(\boldsymbol u^T\boldsymbol v\bigr)^2}{\boldsymbol u^T\Gamma_X\boldsymbol u}.
\end{equation}
For nonsingular $\Gamma_X$, the supremum is
\begin{equation}
\sup_{\boldsymbol u\neq0}\frac{\bigl(\boldsymbol u^T\boldsymbol v\bigr)^2}{\boldsymbol u^T\Gamma_X\boldsymbol u}=\boldsymbol v^T\Gamma_X^{-1}\boldsymbol v,
\end{equation}
and hence
\begin{equation}\label{eq:matrix_area_uncertainty}
\Gamma_Y\succeq\frac{\hbar^2}{4}\Gamma_X^{-1}.
\end{equation}
Note that Eq.~\eqref{eq:vector_Robertson} itself forces $\Gamma_X \succ 0$ for any state with finite second moments.
Taking determinants in Eq.~\eqref{eq:matrix_area_uncertainty} gives
\begin{equation}\label{eq:det_area_monogamy}
\det\Gamma_X\,\det\Gamma_Y\geq\frac{\hbar^4}{16}.
\end{equation}

By construction, $\Gamma_X(t)=\Gamma_{B|A}^{(C)}(t)$.
Moreover, if $J:=\left(\begin{smallmatrix}0&1\\-1&0\end{smallmatrix}\right)$, then $\hat{\boldsymbol Y}=J\bigl(\hat Q_{B|D}^{(C)},\hat\Pi_{B|D}^{(C)}\bigr)^T$, so $\Gamma_Y(t)=J\,\Gamma_{B|D}^{(C)}(t)\,J^T$.
Since $\det J=1$, it follows that $\det\Gamma_Y(t)=\det\Gamma_{B|D}^{(C)}(t)$.
Using the definition~\eqref{eq:AreaDef}, Eq.~\eqref{eq:det_area_monogamy} therefore yields the unconditional covariance-area monogamy relation
\begin{equation}\label{eq:area-monogamy}
{\mathcal A_{B|A}^{(C)}(t)\,\mathcal A_{B|D}^{(C)}(t)\geq\frac{\hbar^2}{4}}.
\end{equation}
Importantly, Eq.~\eqref{eq:area-monogamy} holds for arbitrary states and requires no assumption about either the internal covariances within the two relational pairs or the cross-covariances between the $A$- and $D$-sector observables.

Since 
\begin{equation}
\mathcal A_{B|F}^{(C)}(t)\leq\mathcal M_{B|F}^{(C)}(t),
\end{equation}
for each \(F\in\{A,D\}\), see Eq.~\eqref{eq:M_A_relation}, then the covariance-area bound Eq.~\eqref{eq:area-monogamy} implies the marginal-product relation Eq.~\eqref{eq:M_monogamy_final}. The covariance-area bound is therefore stronger: it constrains the sharpness measures even after correlations within each relational pair have been taken into account.

The result can be given an especially direct trajectory interpretation using the fact that $\mathcal A_{B|F}^{(C)}=\min_{t\in\mathbb R}\mathcal M_{B|F}^{(C)}(t)$ under the opposite-mass free evolution. Then, Eq.~\eqref{eq:area-monogamy} implies
\begin{equation}\label{eq:optimized_monogamy}
\left[\min_{t_A\in\mathbb R}\mathcal M_{B|A}^{(C)}(t_A)\right]\left[\min_{t_D\in\mathbb R}\mathcal M_{B|D}^{(C)}(t_D)\right]\geq\frac{\hbar^2}{4}.
\end{equation}
Thus the monogamy relation cannot be evaded by choosing a different optimal focus time for each candidate frame: even after independent optimization, the two marginal products cannot both approach zero and, therefore, the two trajectories cannot both become arbitrarily sharp.

Finally, the bounds above do not depend on using $C$ as the frame in which the algebra is evaluated, provided that the complete multipartite QRF transformation is applied to both the state and the relational observables. In the present four-system setting, transforming from $C$ to $A$ gives
\begin{equation}
\hat Q_{B|A}^{(C)}\longmapsto\hat x_B^{(A)},\qquad \hat\Pi_{B|A}^{(C)}\longmapsto-\left(\hat p_C^{(A)}+\hat p_D^{(A)}\right),
\end{equation}
while transforming from $C$ to $D$ gives
\begin{equation}
\hat Q_{B|D}^{(C)}\longmapsto\hat x_B^{(D)},\qquad \hat\Pi_{B|D}^{(C)}\longmapsto-\left(\hat p_C^{(D)}+\hat p_A^{(D)}\right).
\end{equation}
Accordingly, the same numerical values of $\mathcal M$ and $\mathcal A$ are obtained in the transformed descriptions, although their explicit operator representations differ from the three-system expressions used in Sec.~\ref{sec:sharpnessDef}. In this precise sense, relational trajectory sharpness is monogamous: arbitrarily sharp localization of the trajectory relative to one negative-mass reference necessarily excludes arbitrarily sharp localization relative to a distinct second reference.

\subsection{Sharp trajectories in composite QRFs}
\label{sec:compQRF}

The monogamy relation established in the previous Subsection constrains the simultaneous sharpness of a probe trajectory relative to two distinct constituent reference systems. It does not, however, rule out the possibility that sharpness is associated with a collective degree of freedom encoded across those systems. We now construct such a collective reference mode from two physical systems \(A\) and \(D\), and show that it can define an arbitrarily sharp relational trajectory for \(B\), even when neither constituent separately provides an arbitrarily sharp reference. 
The collective mode \(F\) is a canonical mode on which the QRF transformation acts; its position \(X_F\) should not be identified with the ordinary center-of-mass position or with a conventional spatial origin of the composite system \(AD\).

Let $\alpha,\beta\in\mathbb{R}$ satisfy $\alpha^2+\beta^2=1$. In frame $C$, define the collective mode
\begin{equation}\label{eq:XF_PF_def}
\hat X_F^{(C)}:=\alpha\hat x_A^{(C)}+\beta\hat x_D^{(C)},\qquad \hat P_F^{(C)}:=\alpha\hat p_A^{(C)}+\beta\hat p_D^{(C)},
\end{equation}
and the orthogonal internal mode
\begin{equation}\label{eq:XR_PR_def}
\hat X_R^{(C)}:=-\beta\hat x_A^{(C)}+\alpha\hat x_D^{(C)},\qquad \hat P_R^{(C)}:=-\beta\hat p_A^{(C)}+\alpha\hat p_D^{(C)}.
\end{equation}
They satisfy $[\hat X_F^{(C)},\hat P_F^{(C)}]=[\hat X_R^{(C)},\hat P_R^{(C)}]=i\hbar$, while all crossed commutators between the $F$ and $R$ modes vanish. Hence the two local modes $(A,D)$ decompose into two commuting canonical modes: a collective frame mode $F$ and an internal mode $R$. The inverse transformation is
\begin{align}
\hat x_A^{(C)}&=\alpha\hat X_F^{(C)}-\beta\hat X_R^{(C)},\qquad \hat x_D^{(C)}=\beta\hat X_F^{(C)}+\alpha\hat X_R^{(C)},\label{eq:xA_xD_inverse}\\
\hat p_A^{(C)}&=\alpha\hat P_F^{(C)}-\beta\hat P_R^{(C)},\qquad \hat p_D^{(C)}=\beta\hat P_F^{(C)}+\alpha\hat P_R^{(C)}.\label{eq:pA_pD_inverse}
\end{align}

The change of variables is implemented by
\begin{equation}\label{eq:UAD_theta}
\hat U_{AD}(\theta)=\exp\left[-\frac{i}{\hbar}\theta\left(\hat x_A^{(C)}\hat p_D^{(C)}-\hat x_D^{(C)}\hat p_A^{(C)}\right)\right],
\end{equation}
with $\alpha=\cos\theta$ and $\beta=\sin\theta$. Its action is $\hat x_A^{(C)}\mapsto\hat X_F^{(C)}$, $\hat p_A^{(C)}\mapsto\hat P_F^{(C)}$, $\hat x_D^{(C)}\mapsto\hat X_R^{(C)}$, and $\hat p_D^{(C)}\mapsto\hat P_R^{(C)}$. Thus $\hat U_{AD}(\theta)$ implements a canonical rotation from the local-mode basis $(A,D)$ to the collective/internal basis $(F,R)$. From this point on, we regard this rotation as defining the mode decomposition $\mathcal H_A\otimes\mathcal H_D\simeq\mathcal H_F\otimes\mathcal H_R$. The QRF transformation introduced below is understood in this representation: the collective mode $F$ plays the role of the new reference system, whereas $R$ remains an ordinary non-frame degree of freedom.

An important consequence of the equal signed masses $m_A=m_D=-m$ is that the free Hamiltonian is invariant in form under this rotation,
$\bigl(\hat p_A^{(C)}\bigr)^2/(2m_A)+\bigl(\hat p_D^{(C)}\bigr)^2/(2m_D)=-\bigl(\hat P_F^{(C)}\bigr)^2/(2m)-\bigl(\hat P_R^{(C)}\bigr)^2/(2m)$. Thus both $F$ and $R$ behave as free modes with effective mass $-m$. In particular, for $m_B=m>0$, the collective mode $F$ has precisely the opposite signed mass to the probe $B$, so that the opposite-mass trajectory construction of the previous Sections applies directly to the pair $(B,F)$.

Once the collective mode $(\hat X_F^{(C)},\hat P_F^{(C)})$ has been identified, we apply to it the instantaneous QRF map $\hat S_x$ used throughout this Section. Since both $B$ and $R$ remain non-frame degrees of freedom, we define
\begin{equation}\label{eq:SxCF_def}
\hat S_x^{(C\to F)}:=\hat{\mathcal P}_{FC}\exp\left[\frac{i}{\hbar}\hat X_F^{(C)}\left(\hat p_B^{(C)}+\hat P_R^{(C)}\right)\right],
\end{equation}
where $\hat{\mathcal P}_{FC}$ denotes the parity-swap between the old frame $C$ and the collective mode $F$. This is the direct analogue of Eq.~\eqref{eq:Sx_def}, with $\hat x_A^{(C)}$ replaced by $\hat X_F^{(C)}$ and with the translation acting on all remaining non-frame degrees of freedom, $B$ and $R$.

In particular,
\begin{equation}\label{eq:Q_BAD_collective}
\hat S_x^{(C\to F)}\left(\hat x_B^{(C)}-\hat X_F^{(C)}\right)\hat S_x^{(C\to F)\dagger}=\hat x_B^{(F)}.
\end{equation}
Here $\hat x_B^{(C)}\mapsto\hat x_B^{(F)}-\hat x_C^{(F)}$ and $\hat X_F^{(C)}\mapsto-\hat x_C^{(F)}$, while the remaining canonical variables transform as $\hat P_F^{(C)}\mapsto-\hat p_C^{(F)}-\hat p_B^{(F)}-\hat P_R^{(F)}$, $\hat p_B^{(C)}\mapsto\hat p_B^{(F)}$, $\hat X_R^{(C)}\mapsto\hat X_R^{(F)}-\hat x_C^{(F)}$, and $\hat P_R^{(C)}\mapsto\hat P_R^{(F)}$.

We now show that there exists a natural class of states for which the collective mode $F$ defines an arbitrarily sharp relational frame for $B$, while neither constituent system $A$ nor $D$ does so individually. The latter statement will rely on a block-diagonal covariance condition between the collective sector $(B,F)$ and the internal mode $R$, which will be stated explicitly below.

For the collective mode, define
\begin{equation}\label{eq:QF_PiF_def}
\hat Q_{B|F}^{(C)}:=\hat x_B^{(C)}-\hat X_F^{(C)},\qquad \hat\Pi_{B|F}^{(C)}:=\hat p_B^{(C)}+\hat P_F^{(C)}.
\end{equation}
These observables commute, $[\hat Q_{B|F}^{(C)},\hat\Pi_{B|F}^{(C)}]=0$, and under the QRF transformation they map as $\hat Q_{B|F}^{(C)}\mapsto\hat x_B^{(F)}$ and $\hat\Pi_{B|F}^{(C)}\mapsto-\bigl(\hat p_C^{(F)}+\hat P_R^{(F)}\bigr)$. Consequently, in the corresponding initial-slice dynamical construction,
\begin{equation}
\hat x_B^{(F)}(t)=\hat x_B^{(F)}(0)-\frac{t}{m}\left[\hat p_C^{(F)}(0)+\hat P_R^{(F)}(0)\right].
\end{equation}
Since the pair $(\hat Q_{B|F}^{(C)},\hat\Pi_{B|F}^{(C)})$ commutes, there is no nonzero uncertainty-relation lower bound on its sharpness. States can therefore be chosen such that both marginal uncertainties become arbitrarily small, 
\begin{equation}\label{eq:composite_sharp_small}
\mathcal M_{B|F}^{(C)}=\Delta\hat Q_{B|F}^{(C)}\,\Delta\hat\Pi_{B|F}^{(C)}\rightarrow0\,,\qquad \mathcal A_{B|F}^{(C)}\rightarrow0\,.
\end{equation}

To compare this collective sharpness with that obtained using either constituent as a reference frame, consider first $A$, with relational pair $\hat Q_{B|A}^{(C)}:=\hat x_B^{(C)}-\hat x_A^{(C)}$ and $\hat\Pi_{B|A}^{(C)}:=\hat p_B^{(C)}+\hat p_A^{(C)}$. Using Eqs.~\eqref{eq:xA_xD_inverse} and \eqref{eq:pA_pD_inverse},
\begin{align}
\hat Q_{B|A}^{(C)}&=\hat Q_{B|F}^{(C)}+(1-\alpha)\hat X_F^{(C)}+\beta\hat X_R^{(C)},\label{eq:QBA_decomp}\\
\hat\Pi_{B|A}^{(C)}&=\hat\Pi_{B|F}^{(C)}+(\alpha-1)\hat P_F^{(C)}-\beta\hat P_R^{(C)}.\label{eq:PiBA_decomp}
\end{align}
Similarly, for $D$,
\begin{align}
\hat Q_{B|D}^{(C)}&=\hat Q_{B|F}^{(C)}+(1-\beta)\hat X_F^{(C)}-\alpha\hat X_R^{(C)},\label{eq:QBD_decomp}\\
\hat\Pi_{B|D}^{(C)}&=\hat\Pi_{B|F}^{(C)}+(\beta-1)\hat P_F^{(C)}+\alpha\hat P_R^{(C)}.\label{eq:PiBD_decomp}
\end{align}
Thus the local relational pairs associated with $A$ and $D$ are not identical to the collective pair associated with $F$: each contains a contribution from the internal mode $R$, together with residual collective-sector terms.

To obtain explicit local lower bounds, assume that the state is covariance-block-diagonal between the internal mode $R$ and the collective sector $(B,F)$, namely that all covariances between $\hat X_R^{(C)},\hat P_R^{(C)}$ and operators constructed from the $B$ and $F$ modes vanish. Under this assumption,
\begin{align}
\Delta^2\hat Q_{B|A}^{(C)}&=\Delta^2\left[\hat Q_{B|F}^{(C)}+(1-\alpha)\hat X_F^{(C)}\right]+\beta^2\Delta^2\hat X_R^{(C)},\label{eq:VarQBA_bound}\\
\Delta^2\hat\Pi_{B|A}^{(C)}&=\Delta^2\left[\hat\Pi_{B|F}^{(C)}+(\alpha-1)\hat P_F^{(C)}\right]+\beta^2\Delta^2\hat P_R^{(C)}.\label{eq:VarPiBA_bound}
\end{align}
Hence $\Delta\hat Q_{B|A}^{(C)}\geq|\beta|\Delta\hat X_R^{(C)}$ and $\Delta\hat\Pi_{B|A}^{(C)}\geq|\beta|\Delta\hat P_R^{(C)}$, which gives
\begin{equation}\label{eq:MBA_lower_bound}
\mathcal M_{B|A}^{(C)}\geq\beta^2\Delta\hat X_R^{(C)}\,\Delta\hat P_R^{(C)}\geq\beta^2\frac{\hbar}{2}.
\end{equation}
Analogously,
\begin{equation}\label{eq:MBD_lower_bound}
\mathcal M_{B|D}^{(C)}\geq\alpha^2\Delta\hat X_R^{(C)}\,\Delta\hat P_R^{(C)}\geq\alpha^2\frac{\hbar}{2}.
\end{equation}
Thus, under the stated covariance assumption, the internal canonical mode prevents either constituent reference frame from becoming arbitrarily sharp whenever both $\alpha$ and $\beta$ are nonzero.

The same conclusion can be strengthened using the covariance-area measure introduced in Eq.~\eqref{eq:AreaDef}. Define $D:=\mathrm{diag}(1,-1)$ and $\hat{\boldsymbol R}_R^{(C)}:=(\hat X_R^{(C)},\hat P_R^{(C)})^{\mathsf T}$, and let $\Gamma_R^{(C)}$ be its covariance matrix. From Eqs.~\eqref{eq:QBA_decomp} and \eqref{eq:PiBA_decomp},
\begin{equation}
\begin{pmatrix}\hat Q_{B|A}^{(C)}\\ \hat\Pi_{B|A}^{(C)}\end{pmatrix}
=
\begin{pmatrix}\hat Q_{B|F}^{(C)}+(1-\alpha)\hat X_F^{(C)}\\ \hat\Pi_{B|F}^{(C)}+(\alpha-1)\hat P_F^{(C)}\end{pmatrix}
+\beta D\hat{\boldsymbol R}_R^{(C)}.
\end{equation}
Under the same block-diagonal covariance assumption, $\Gamma_{B|A}^{(C)}=\Gamma_{A,\mathrm{coll}}^{(C)}+\beta^2D\Gamma_R^{(C)}D\succeq\beta^2D\Gamma_R^{(C)}D$, where $\Gamma_{A,\mathrm{coll}}^{(C)}\succeq0$ is the covariance matrix of the first vector on the right-hand side. Since $(\hat X_R^{(C)},\hat P_R^{(C)})$ is canonical, the Robertson--Schr\"odinger relation gives $\det\Gamma_R^{(C)}\geq\hbar^2/4$. Taking determinants therefore yields
\begin{equation}\label{eq:ABA_lower_bound}
\mathcal A_{B|A}^{(C)}\geq\beta^2\sqrt{\det\Gamma_R^{(C)}}\geq\beta^2\frac{\hbar}{2}.
\end{equation}
Exactly the same argument applied to Eqs.~\eqref{eq:QBD_decomp} and \eqref{eq:PiBD_decomp} gives
\begin{equation}\label{eq:ABD_lower_bound}
\mathcal A_{B|D}^{(C)}\geq\alpha^2\sqrt{\det\Gamma_R^{(C)}}\geq\alpha^2\frac{\hbar}{2}.
\end{equation}
These bounds are stronger than Eqs.~\eqref{eq:MBA_lower_bound} and \eqref{eq:MBD_lower_bound}, since $\mathcal A_{B|F}^{(C)}\leq\mathcal M_{B|F}^{(C)}$ for every relational pair. 

For the balanced collective mode $\alpha=\beta=1/\sqrt{2}$, the local bounds become
\begin{equation}\label{eq:local_bounds_summary}
\mathcal M_{B|A}^{(C)}\geq\frac{\hbar}{4},\qquad \mathcal M_{B|D}^{(C)}\geq\frac{\hbar}{4},
\end{equation}
and, more strongly,
\begin{equation}\label{eq:local_area_bounds_summary}
\mathcal A_{B|A}^{(C)}\geq\frac{\hbar}{4},\qquad \mathcal A_{B|D}^{(C)}\geq\frac{\hbar}{4}.
\end{equation}
At the same time, since the collective relational pair commutes, states can be chosen such that $\Delta\hat Q_{B|F}^{(C)}\rightarrow0$ and $\Delta\hat\Pi_{B|F}^{(C)}\rightarrow0$, and consequently $\mathcal M_{B|F}^{(C)}\rightarrow0$ and $\mathcal A_{B|F}^{(C)}\rightarrow0$. Thus the trajectory of $B$ can be made arbitrarily sharp relative to the collective mode $F$ in the stronger sense that both trajectory-defining observables can be made simultaneously arbitrarily precise, even though, within the stated class of states, the corresponding local sharpness measures remain bounded away from zero for either constituent $A$ or $D$ separately.

This does not violate the monogamy relations derived in the previous Subsection. Those relations constrain simultaneous sharpness with respect to the two distinct local reference systems $A$ and $D$. Here the sharp trajectory is instead defined with respect to a different reference degree of freedom, namely the collective mode $F$ encoded jointly across $A$ and $D$. The result therefore shows that relational trajectory sharpness can be encoded in a distributed collective degree of freedom even when it is unavailable to either constituent reference frame individually.

%%%%%%%%%%%%%%%%%%%%%%%%%%%%%%%%%%%%%%%%%%%%%%%%%%%%%%%%%%%%%%%%%%%%%

\section{Conclusions}

We have shown that trajectory-sharpening through EPR-type correlations admits a natural interpretation in terms of quantum reference frame (QRF) transformations. 
In a minimal toy model of two free particles with opposite-sign masses, the commuting observables that determine the relational trajectory of a probe map, under the instantaneous position-QRF transformation,  to operators local to distinct subsystems in the frame of the negative-mass reference.  We have further established that the dynamical QRF transformation reproduces the same relational trajectory. 
The two descriptions distribute the time dependence differently: in the initial-slice construction, the trajectory is encoded in the evolution generated by a transformed, non-free Hamiltonian, whereas in the extended-symmetry construction, the Hamiltonian retains its free-particle form and the transformed relational position becomes explicitly time dependent.
This equivalence clarifies how a trajectory-sharpening effect familiar from quantum metrology can be understood within the operational framework of QRFs.

Building on this correspondence, we have introduced two QRF-invariant measures of trajectory sharpness. The marginal uncertainty product $\mathcal M_{B|F}$ provides an instantaneous measure of the marginal uncertainty area of the trajectory-defining observable, while the covariance-area parameter $\mathcal A_{B|F}$ additionally accounts for correlations between them. We showed that both quantities are invariant under QRF transformations, and that $\mathcal A_{B|F}$ is exactly conserved under the opposite-mass free evolution. Moreover, $\mathcal A_{B|F}$ equals the minimum value of $\mathcal M_{B|F}(t)$ attained along the free relational trajectory.

These measures expose a monogamy relation for relational trajectory sharpness. Rendering the trajectory of a probe system $B$ sharp relative to one negative-mass frame $A$ necessarily limits how sharp the same trajectory can be made relative to another distinct frame $D$. In particular, we have obtained the unconditional bounds
\begin{equation}
\mathcal A_{B|A}^{(C)}\,\mathcal A_{B|D}^{(C)}\geq\frac{\hbar^2}{4},
\qquad
\mathcal M_{B|A}^{(C)}\,\mathcal M_{B|D}^{(C)}\geq\frac{\hbar^2}{4}.
\end{equation}
Since $\mathcal A_{B|F}$ is the minimum marginal sharpness attained along the corresponding free trajectory, the first relation shows that the trade-off persists even when each trajectory is evaluated at its own optimal focus time. In this precise sense, trajectory sharpness with respect to distinct local negative-mass QRFs constitutes a non-shareable resource.

Finally, we have shown that this monogamy constraint does not prevent trajectory sharpness from residing in a collective reference degree of freedom. An effective distributed QRF can be constructed from a collective canonical mode encoded jointly in two physical systems, and the trajectory of $B$ can be made arbitrarily sharp relative to this collective mode, with both trajectory-defining uncertainties simultaneously approaching zero, even when, for the class of states considered, neither constituent system alone can provide an arbitrarily sharp reference. 
Under the stated block-diagonal covariance condition, the residual lower bounds on the constituent-frame sharpness can be traced to the conjugate fluctuations of the internal mode orthogonal to the collective frame mode. This provides a transparent algebraic mechanism by which relational trajectory sharpness can be encoded in a distributed degree of freedom rather than localized in either constituent separately.

Beyond their conceptual interest, these results suggest that the sharpening of a system's trajectory via EPR correlations with a negative-mass reference, already demonstrated experimentally with atomic spin ensembles and mechanical oscillators, can be understood and further generalized within the language of quantum reference frames. 
This opens the door to identifying new, possibly genuinely QRF-specific predictions that can be tested in optomechanical or atomic-ensemble platforms, where the interplay between QRF covariance, relational dynamics, and trajectory sharpness may reveal further constraints of operational and foundational significance.

\begin{acknowledgements}

MF was supported by the Swiss National Science Foundation Ambizione Grant No. 208886, and by The Branco Weiss Fellowship -- Society in Science, administered by the ETH Z\"{u}rich. 
The authors would like to acknowledge networking support by the COST Action CA23130 ``BridgeQG: Bridging high and low energies in search of quantum gravity''.
\end{acknowledgements}

%%%%%%%%%%%%%%%%%%%%%%%%%%%%%%%%%%%%%%%%%%%%%%%%%%%%%%%%%%%%%%%%%%%%%

\appendix
\section{Boost transformation for negative-mass QRF }
\label{sec:supp-boost}

Although we have seen that the position QRF transformation does not depend on the mass of the systems, the situation is different for boosts. The latter is defined as \cite{Ballesteros:2020lgl}
\begin{equation}\label{eq:defSb}
\hat S_b^{(C\to A)} = e^{-\frac{i}{\hbar}\hat Q_C^{(A)} t}\, \hat{\mathcal{P}}_{AC}\, e^{\frac{i}{\hbar}\log \frac{m_C}{m_A} \Dac}\, e^{\frac{i}{\hbar}\hat K_{AB}^{(C)}}\, e^{\frac{i}{\hbar}\hat Q_A^{(C)} t} ,
\end{equation}
where $\Dac=\left(\xac\pac + \pac\xac \right)/2$, corresponding to the dilation generator on the phase space of $A$, and $\hat{K}_{AB}^{(C)} = \pac \hat G_B^{(C)} / m_A$, with $\hat G_B^{(C)} = \pbc\, t - m_B\,\xbc$ the generator of the standard Galilean boost on particle $B$. Equation~\eqref{eq:defSb} contains the generalized parity-swap term
\begin{equation}
\hat{\mathcal P}^{(v)}_{AC} = \hat{\mathcal P}_{AC} \exp\left[ \frac{i}{\hbar}\log \frac{m_C}{m_A}\, \Dac \right] .
\label{eq:positive_mass_generalized_swap}
\end{equation}
The role of the extra exponential factor is to rescale position and momentum before the parity-swap. Indeed, for any real parameter \(\alpha\),
\begin{subequations} \label{eq:dilation_action}
\begin{align}
e^{\frac{i}{\hbar}\alpha \Dac}\, \xac \,e^{-\frac{i}{\hbar}\alpha \Dac} &=e^{\alpha} \xac, \\
e^{\frac{i}{\hbar}\alpha \Dac}\, \pac \,e^{-\frac{i}{\hbar}\alpha \Dac} &=e^{-\alpha} \pac.
\end{align}
\end{subequations}
Thus \(\exp[\frac{i}{\hbar}\alpha\Dac]\) is a single-mode squeezing or phase-space dilation. This factor is required because for a boost transformation one must transform the velocity of $A$ relative to $C$ into minus the velocity of $C$ relative to $A$ according to
\begin{equation}\label{eq:VelSwap}
    \dfrac{\pac}{m_A} \rightarrow - \dfrac{\pca}{m_C} \;.
\end{equation}
A bare parity-swap \(\hat{\mathcal P}_{AC}\) cannot achieve this if \( m_A\neq m_C\), because it interchanges canonical operators but does not rescale them by the mass ratio.

The expression \eqref{eq:positive_mass_generalized_swap} is well defined as a unitary operator when \( m_C/ m_A>0\), because then the coefficient $\log \frac{ m_C}{ m_A}$ is real. However, if the old and new frames have opposite-sign masses, $\log \frac{ m_C}{ m_A}$ is no longer real, so the exponent in \eqref{eq:positive_mass_generalized_swap} is not anti-Hermitian. Therefore the positive-mass formula cannot be used naively when \( m_A m_C<0\). This is the precise sense in which the boost transformation requires a modification for a negative-mass QRF.

To construct a consistent negative-mass extension of the generalized parity-swap, we separate the magnitude of the mass ratio from its sign. Define
\begin{equation}
\lambda_{CA}:=\left|\frac{ m_C}{ m_A}\right|>0, \qquad \sigma_{CA}:=\mathrm{sgn}\left(\frac{ m_C}{ m_A}\right)\in\{+1,-1\}.
\label{eq:lambda_sigma_def}
\end{equation}
We also introduce the parity operator \(\hat \Pi_C\) acting on system $C$ as
\begin{subequations}
\begin{align}
\hat \Pi_C \xca \hat \Pi_C^\dagger &= -\xca, \\
\hat \Pi_C \pca \hat \Pi_C^\dagger &= -\pca.
\end{align}
\label{eq:parity_C_def}
\end{subequations}
Then we define the \emph{signed generalized parity-swap}
\begin{equation}
\hat{\mathcal P}^{(v,\mathrm{sgn})}_{AC} := \hat \Pi_C^{\frac{1-\sigma_{CA}}{2}}\, \hat{\mathcal P}_{AC}\, \exp\left[ \frac{i}{\hbar}\log\lambda_{CA}\, \Dac \right].
\label{eq:signed_generalized_swap}
\end{equation}
This operator is unitary for all nonzero signed masses \( m_A, m_C\), because \(\lambda_{CA}>0\) and therefore \(\log\lambda_{CA}\in\mathbb{R}\). This allows us to identify $\alpha=\log\lambda_{CA}$ in Eq.~\eqref{eq:dilation_action}, which results in
\begin{subequations}
\begin{align}
\exp\left[ \frac{i}{\hbar}\log\lambda_{CA}\, \Dac \right] \xac \exp\left[ -\frac{i}{\hbar}\log\lambda_{CA}\, \Dac \right] &= \lambda_{CA}\,\xac, \\
\exp\left[ \frac{i}{\hbar}\log\lambda_{CA}\, \Dac \right] \pac \exp\left[ -\frac{i}{\hbar}\log\lambda_{CA}\, \Dac \right] &= \frac{1}{\lambda_{CA}}\,\pac.
\end{align}
\end{subequations}
Then the ordinary parity-swap $\hat{\mathcal P}_{AC}$ gives
\begin{equation}
\hat{\mathcal P}_{AC}\xac\hat{\mathcal P}_{AC}^\dagger=-\xca, \qquad \hat{\mathcal P}_{AC}\pac\hat{\mathcal P}_{AC}^\dagger=-\pca.
\end{equation}
Finally, if \(\sigma_{CA}=-1\), the extra parity operator \(\hat \Pi_C\) flips the sign once more. Altogether one obtains
\begin{subequations}
\begin{align}
\hat{\mathcal P}^{(v,\mathrm{sgn})}_{AC}\,\xac\,\hat{\mathcal P}^{(v,\mathrm{sgn})\dagger}_{AC} &= -\sigma_{CA}\lambda_{CA}\,\xca , \\
\hat{\mathcal P}^{(v,\mathrm{sgn})}_{AC}\,\pac\,\hat{\mathcal P}^{(v,\mathrm{sgn})\dagger}_{AC} &= -\frac{\sigma_{CA}}{\lambda_{CA}}\,\pca .
\end{align}
\label{eq:signed_swap_action_xp}
\end{subequations}
Using \(\sigma_{CA}\lambda_{CA}= m_C/m_A\) and \(\sigma_{CA}/\lambda_{CA}= m_A/m_C\), the momentum transformation gives
\begin{equation}
\hat{\mathcal P}^{(v,\mathrm{sgn})}_{AC}\, \frac{\pac}{ m_A}\, \hat{\mathcal P}^{(v,\mathrm{sgn})\dagger}_{AC} = -\frac{\pca}{m_C},
\label{eq:velocity_map_signed_final}
\end{equation}
which is exactly the desired velocity inversion Eq.~\eqref{eq:VelSwap}.

We can now write the boost-type QRF transformation for a signed-mass reference system $A$ as the unitary operator
\begin{equation}\label{eq:modSb}
\hat S_b^{(C\to A)} = e^{-\frac{i}{\hbar}\hat Q_C^{(A)} t}\, \hat \Pi_C^{\frac{1-\sigma_{CA}}{2}} \hat{\mathcal{P}}_{AC}\, e^{\frac{i}{\hbar}\log \lambda_{CA} \Dac}\, e^{\frac{i}{\hbar}\hat K_{AB}^{(C)}}\, e^{\frac{i}{\hbar}\hat Q_A^{(C)} t} .
\end{equation}
Note that the operator $\hat\Pi_C$ is crucial to get the desired Eq.~\eqref{eq:VelSwap} even when $m_A<0$. In fact, just forcing the argument inside the logarithm to be positive would lead to $\pac/m_A \rightarrow + \pca/m_C$.

By the same Baker--Campbell--Hausdorff manipulations as in the positive-mass case, one finds that the action of Eq.~\eqref{eq:modSb} on the observables of $A$, $B$ is
\begin{subequations}
\begin{align}
\hat S_b\,\xac\, \hat S_b^{\dagger} &= - \dfrac{m_C}{m_A} \xca + \left( \dfrac{1}{m_A} - \dfrac{1}{m_C} \right) \pca t + \left( \dfrac{\pba}{m_A} t - \dfrac{m_B}{m_A} \xba \right) , \\
\hat S_b\,\pac\, \hat S_b^{\dagger} &= -\frac{m_A}{m_C} \pca ,\\
\hat S_b\,\xbc \,\hat S_b^{\dagger} &= \xba -\frac{\pca}{m_C}t, \\
\hat S_b\,\pbc\,\hat S_b^{\dagger} &= \pba-\frac{ m_B}{ m_C}\pca .
\end{align}\label{eq:Sb_signed_transf}
\end{subequations}
Note that, although these equations have the same algebraic form as the positive-mass formulas presented in Ref.~\cite{Ballesteros:2020lgl}, they now hold for all nonzero signed masses because the boost operator in Eq.~\eqref{eq:modSb} now admits negative $m_C/m_A$.

\bibliography{biblio}

\end{document}